\documentclass[12pt,preprint]{aastex}

\newcommand{\simlt}{\lower.5ex\hbox{$\; \buildrel < \over \sim \;$}}

\slugcomment{draft date: \today}

\begin{document}

\title{Properties of protostars in the 
Elephant Trunk globule IC 1396A}

\author{William T. Reach\altaffilmark{1,2},
Dohy Faied\altaffilmark{2},                
Jeonghee Rho\altaffilmark{2},               
Adwin Boogert\altaffilmark{1},             
Achim Tappe\altaffilmark{2,3},             
Thomas H. Jarrett\altaffilmark{1,2}
Patrick Morris\altaffilmark{1},
Laurent Cambr\'esy\altaffilmark{4},        
Francesco Palla\altaffilmark{5},
Riccardo Valdettaro\altaffilmark{5}
}

\altaffiltext{1}{Infrared Processing and Analysis Center,MS 100-22,
California Institute of Technology,
Pasadena, CA 91125}

\altaffiltext{2}{{\it Spitzer} Science Center, MS 220-6, 
California Institute of Technology,
Pasadena, CA 91125}

\altaffiltext{3}{Harvard-Smithsonian Center for Astrophysics, 60 Garden St., Cambridge, MA 02138}

\altaffiltext{4}{Observatoire de Strasbourg, Strasbourg, France}

\altaffiltext{5}{NAF Osservatorio Astrofisico di Arcetri, Largo Enrico Fermi 5, 50125 Florence, Italy}

\email{reach@ipac.caltech.edu}

\begin{abstract}

Extremely red objects, identified in the early {\it Spitzer}
Space Telescope observations of the bright-rimmed globule IC 1396A
and photometrically classified as Class I protostars Class II T Tauri stars
based on their mid-infrared colors, were observed spectroscopically
at 5.5--38 $\mu$m ({\it Spitzer} InfraRed Spectrograph), 
at the 22 GHz water maser frequency (NRAO Green Bank Telescope), 
and in the optical (Palomar Hale 5-m), to
confirm their nature and further elucidate their properties.
The sources photometrically identified as Class I,
including IC1396A:$\alpha$, $\gamma$, $\delta$, $\epsilon$, and $\zeta$, are confirmed as
objects dominated by accretion luminosity from dense envelopes, with 
accretion rates 1--10$\times 10^{-6} M_{\odot}$~yr$^{-1}$ and present stellar
masses 0.1--2 $M_{\odot}$.
The Class I sources have extremely
red continua, still rising at 38 $\mu$m, with a deep silicate 
absorption at 9--11 $\mu$m, weaker silicate absorption around 18 $\mu$m,
and weak ice features including CO$_2$ at 15.2 $\mu$m and H$_2$O at
6 $\mu$m. 
The ice/silicate absorption ratio in the envelope is exceptionally low
for the IC 1396A protostars, compared to those in nearby star-forming regions,
suggesting the envelope chemistry is altered by the radiation field or globule
pressure.
Only one 22 GHz water maser 
was detected in IC 1396A; it is coincident with a faint mid-infrared source, 
offset from near the luminous Class I protostar IC 1396A$\gamma$.
The maser source, IC 1396A$\gamma_{b}$, has luminosity $<0.1 L_{\odot}$, the first
H$_{2}$O maser from such a low-luminosity object.
Two near-infrared H$_{2}$ knots on opposite sides of IC 1396A:$\gamma$ reveal a jet,
with axis clearly distinct from the H$_{2}$O maser of IC 1396A:$\gamma_{b}$.

The objects photometrically classified as Class II,
including IC1396A:$\beta$, $\theta$, 2MASSJ 21364964+5722270,
2MASSJ 21362507+5727502, 
LkH$\alpha$ 349c, Tr 37 11-2146 and Tr 37 11-2037,
are confirmed as stars with warm, luminous disks, with a
silicate emission feature at 9--11 $\mu$m, and bright H$\alpha$
emission, so they are young, disk-bearing, classical T Tauri stars.
The disk properties change significantly with source luminosity:
low-mass (G--K) stars have prominent 9--11 emission features due
to amorphous silicates while higher-mass (A--F) stars have weaker
features requiring abundant crystalline silicates.
A mineralogical model that fits the wide and low-amplitude silicate
feature of IC1396A:$\theta$ requires small grains of crystalline olivine (11.3 $\mu$m peak)
and another material to to explain the its 9.1 $\mu$m peak;
reasonable fits are obtained with a phyllosilicate, quartz, or
relatively large ($>10$ $\mu$m) amorphous olivine grains.

The distribution of Class I sources is concentrated within the molecular globule, while
the  Class II sources are more widely scattered. Combined with the spectral results,
this suggests two
phases of star formation, the first (4 Myr ago) leading
to the widespread Class II sources and the central O star of IC 1396,
and the second ($< 1$ Myr ago) 
occurring  within the globule. The recent phase was
likely triggered by the wind and radiation of the central O star 
of the IC 1396 \ion{H}{2} region.

\end{abstract}


\section{Introduction}

{\it Spitzer} has opened widely the mid-infrared (3.6--24 $\mu$m) window for
studies of star formation.
In the earliest observations with the observatory,
as in many subsequent ones,
previously unknown, mid-infrared-bright sources have been discovered in
star-forming regions.
In this paper we follow up \citep[][Paper I]{reachIC} on the mid-infrared sources in the bright-rimmed 
globule IC 1396A, 
known as the Elephant Trunk Nebula, a dense globule in the large, nearby (750 pc) 
\ion{H}{2} region IC 1396, which is is excited by the 4 Myr-old O6 star HD 206267 \citep[see][]{weikard}.
Based on comparison to measured colors to those of 
young stellar objects in Taurus \citep{kenyonhartmann}, we suggested
10 sources may be the very early Class I stage, which is thought to only last $10^{5}$ yr
\citep{spbook}, indicating that the globule is a site of very recent star formation.

Detecting new protostars in a wide variety of star forming regions
is an important expansion of previous work that by necessity has
concentrated on the very closest star forming regions such as
Taurus and Orion. While these nearby regions provide excellent
opportunities to observe the widest range of star formation (from
brown dwarfs to massive stars), they are necessarily parochial.
Contrary to a long-standing common belief that the Sun formed in a
quiescent region like Taurus, meteoritic evidence indicates the Sun
likely formed near a high-mass star, whose \ion{H}{2} region, winds, and
supernova explosion influence the evolution and composition of
the protoplanetary disk \citep{meteorite,tachibana}.
To search for present-day star-forming regions similar to
those in which the Sun formed, we must consider massive star forming regions.
In such regions, stars can be formed by radiative driven implosion of
moderately dense cores \citep{lefloch},
and planet-building disks are exposed to strong ultraviolet 
radiation \citep{proplyd}. A major goal of future star formation studies
should be to determine the effects of massive stars both in triggering
star formation and shaping disks around young stellar objects.

In this paper, we present follow-up observations to examine the
nature of the sources in IC 1396A in more detail. 
New {\it Spitzer} mid-infrared 
spectra demonstrate that the photometrically extremely red objects are indeed
Class I protostars, with dense, highly-optically-thick envelopes
shrouding a moderately luminous core inside a dust photosphere.
We show that at least one of the extremely red objects, IC 1396 A:$\gamma$,
powers a molecular outflow with H$_2$O maser emission, characteristic
of a very young ($\sim 10^{4}$ yr) Class 0 protostars \citep{furuya}.
The photometrically-classified Class II objects are compared to 
classical T Tauri stars, based on H$\alpha$ emission and 
mid-infrared disk spectra.

\section{Observations\label{obssec}}

\subsection{Spitzer Space Telescope}
Mid-infrared spectra were obtained using the 
InfraRed Spectrograph \citep[IRS;][]{houck} in 2004 and 2005 November.
Table~\ref{opttab} lists the sources and IRS observing modes.
The 2004 observations was adversely affected
by poor pointing; the 2005 observations repeated
those of the first set that had offsets between actual and expected pointing
larger than half the slit width in the cross-slit direction.
Spectra were taken with both the short- and long-wavelength,
low-resolution mode for all sources. The brightest sources
were also observed with the high-resolution mode.
Observations were performed in groups (`clusters'), and
a single bright, isolated mid-infrared source (whose coordinates
were measured from the IRAC images taken in 2003 December) was
used as a peak-up source for each group. 
Mid-infrared source brightnesses and positions were verified using the IRAC and MIPS
images from Paper I. The magnitudes are reported in Table~\ref{magtab}.
With respect to the values in Paper I, significant changes include the following.
Source IC 1396A:$\lambda$ is significantly fainter
in the IRAC 3.6 and 4.5 $\mu$m channels, because in Paper I it was 
confused with another nearby source that is brighter in those channels
(but far fainter at 24 $\mu$m). The changed colors move the source in
the color-color diagram (Fig. 3 of Paper I) from an unlabeled region
into the Class 1 region. The position of source IC 1396A:$\kappa$ was
corrected from the typographical error (copied declination of $\lambda$).

For the basic reductions, the on-slit nods were subtracted from
each other, and spectra were extracted using SMART \citep{smartref}.
In many cases, the nod-subtracted spectra were contaminated by other sources or
bright structures in the nebula; for these, a local background subtraction was performed
on each nodded image.
The extracted spectra from each nod position were averaged, and their uncertainties
combined using the root-sum-square of each extraction's uncertainty.

For IC 1396A:$\alpha$, $\gamma$, $\delta$, and $\epsilon$, the low-resolution
($R=\lambda/\Delta\lambda\sim$90) 2-dimensional Basic Calibrated Data (BCD) spectral 
images produced by the SSC pipeline versions S15.3.0 and S17.2.0 were used as
the starting point of a customized reduction. First, the spectral and
spatial dimensions were orthogonalized, and then the 2-dimensional
images of the two nodding positions were subtracted in order to remove
extended background emission.  In some cases the background emission
is highly structured or multiple sources are present in the slit, and
nodding pairs could not be subtracted.  Here, the background observed
in nearby telescope pointings was subtracted and any residual emission
was removed using the background determined from neighboring columns
in the source extraction process. Some background emission lines
(especially at 6.2, 17.0, and 18.7 $\mu$m) were not removed completely
or were over-corrected for, which was taken into account in the
interpretation of the YSO spectra.  The background subtraction process
also removes most of the effects of pixels with deviating dark
currents (``hot pixels''). Any remaining bad pixels were replaced by
values from a two-dimensional interpolation.  Subsequently a fixed
width (4 pixels) extraction was performed and the 1-dimensional
spectra were then averaged. Then the spectra were divided by spectra
of the standard star HR 2194 reduced in the same way in order to
correct for wavelength-dependent slit losses. Spectral features in HR
2194 (A0~V) were divided out using the photospheric model of Decin et
al. (2004). Some residual standing wave structure with a period of
$\sim$1 $\mu$m observed above 20 $\mu$m was not corrected for.
Finally, the Spitzer/IRS modules and ground-based spectra were
multiplied along the flux scale in order to match Spitzer/IRAC 3.6 and
8.0 $\mu$m fluxes, using the appropriate filter profiles, and to
provide smooth transitions between the IRS modules.

\subsection{Keck/NIRSPEC L-band Spectra}

The Spitzer spectra of IC 1396A $\alpha$ and $\gamma$ are complemented
by ground-based L-band spectra obtained with the NIRSPEC spectrometer
at the Keck II telescope at Mauna Kea \citep{mcl98} at a resolving
power of $R=\lambda/\Delta\lambda$=2000. This provides an independent
measure of the solid H$_2$O column density through the strong O-H
stretching mode at 3.0 $\mu$m. The data were obtained on UT 24 July
2004 in two separate grating settings covering the 2.70-3.54 and
3.34-4.19 $\mu$m spectral ranges.  The telescope was nodded on the sky
in an ABBA pattern for 12 and 8 minutes for the short and
long-wavelength settings respectively.  The spectra were reduced in a
way standard for ground-based long-slit spectra, using the nearby main
sequence star HR 8208 (F0V) for telluric line correction.  The final
signal-to-noise values are 30 and 50 for the short and long wavelength
sections, respectively.  Finally, the ground-based spectra were
multiplied along the flux scale in order to match the Spitzer/IRAC 3.6
$\mu$m photometry, using the appropriate filter profile.

\subsection{Palomar Hale Telescope: Optical spectra}

Visible spectra were obtained using the Double Spectrograph \citep{oke}
in August 2006. Visible counterparts were identified for the Class II
sources using finder charts from the Palomar Observatory Sky Survey (POSS),
2nd-generation Digitized Sky Survey; these sources were identified
easily and placed in the long slit with $1''$ width. 
For objects not visible on the POSS, offsets were performed from
a nearby star, using the coordinates from the IRAC images relative
to the star in the POSS.
All spectra were obtained in the low-resolution mode, with a
resolving power 2100 (2500) in the blue (red) portion of the spectra.
A dichroic with transition at 5500 \AA\ was used
to separate light to the blue and red gratings and cameras. 
The blue spectrum was centered at 4400 \AA, with useful
range 2700-5500 \AA.
The red spectrum was centered at 6800 \AA, with useful
range 5500-7900 \AA. The spectra were analyzed using IRAF and then normalized in IDL. 
For objects not visible on the POSS, their spectra were extracted using a reference position on 
the slit from a bright sources. 

\def\extra{
From \citet{osterbrock}, the ratio of Balmer lines for recombining gas is 
$j($H$\beta)$/$j($H$\alpha)$= 0.350 for Case A and 0.348 for Case B.
}

\subsection{Green Bank Telescope}

Spectroscopic observations to search for H$_2$O masers from
fourteen mid-infrared sources in IC 1396A were performed using the 
National Radio Astronomy Observatory 
(NRAO)\footnote{The National Radio Astronomy Observatory is a facility of the
National Science Foundation operated under cooperative agreement by Associated
Universities, Inc.}
100-m Robert C. Byrd Green Bank Telescope.
The targets were observed in multiple sessions, to allow for the possible
strong variability of protostellar masers on month timescales. 
The observing sessions were
on 2006 Aug 7 and 2006 Sep 27.

The 18-22 GHz receiver was tuned to 22.23508 GHz, and the spectroscopic
observations were taken in two methods. 
First, each source was observed by nodding between the two beams of the receiver;
this efficient technique always has the source within one of the beams, but it
only works the `off' beam located $3^{}\prime$ in azimuth is free of emission.
Second, each source was observed by position-switching to a blank (in radio and
mid-infrared emission) field. 
Both the nods and position-switched
spectra were analyzed using GBTIDL. The well-known maser source S Per
was observed to validate the observing technique, and it was cleanly
detected with an antenna temperature 140 K, consistent with the
previously-measured
gain of 1.8 K/Jy and the previously-measured 
S Per flux of 76 Jy \citep{vlemmings}.

In all cases, the nodded and position-switched spectra agreed, and for
all spectra except those near IC 1396A:$\gamma$ the spectra are consistent
with noise.
Upper limits for undetected sources were derived as follows.
Each 1-min-per-beam nodded spectrum yielded an rms noise of 0.24 K antenna
temperature with 0.01 km~s$^{-1}$ autocorrelator bins after hanning smoothing.
Rebinning to 0.1 (1) km~s$^{-1}$ bins, the rms antenna temperature was 
0.11 (0.04) K. 
Therefore,
we place a 3-$\sigma$ upper limit on the H$_2$O masers for all observed
sources (except detections noted below) of
0.18 Jy peak (in 0.1 km~s$^{-1}$ bins) and 0.07 Jy~km~s$^{-1}$ 
flux (in  1 km~s$^{-1}$ bins).
For sources IC 1396A:$\gamma$, $\alpha$, and $\delta$,
observations were 8 times longer yielding an rms of 0.028 K, 
and a 3-$\sigma$ upper limits of 0.05 Jy peak
and 0.02 Jy~km~s$^{-1}$ flux.

\begin{figure}[th]
\epsscale{.5}
\plotone{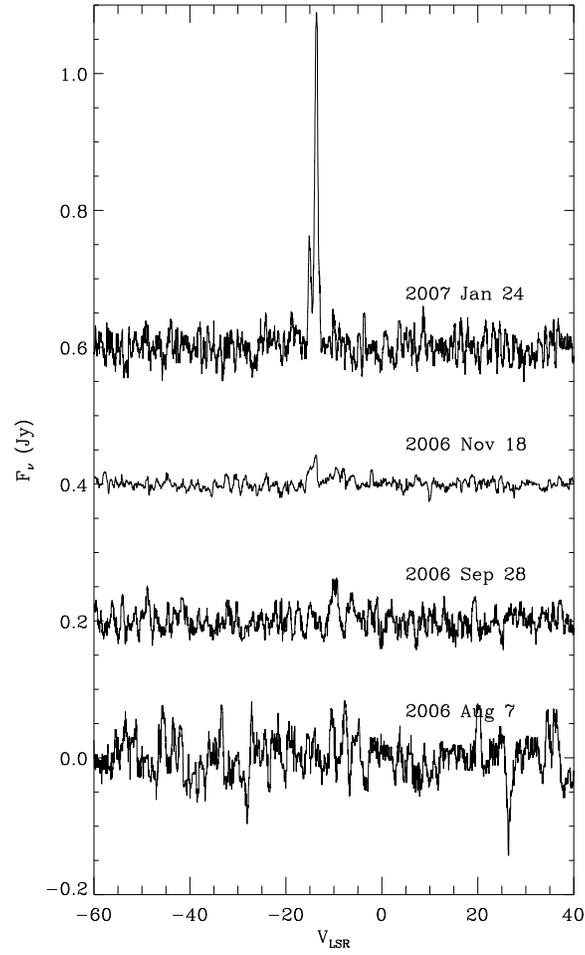}
\epsscale{1}
\figcaption[plotgammany]{GBT 
22 GHz H$_2$O maser spectra of IC1396A:$\gamma$ on 4 dates (as labeled). The
spectra are offset vertically by 0.2 Jy on each successive date, for clarity.
\label{gamwater}}
\end{figure}

Figure~\ref{gamwater} shows the H$_2$O maser spectra of IC1396A:$\gamma$.
This source had been previously detected by \citet{valdettaro05} using the
Medicina telescope with 1.9$'$ beam. 
Relative to the peak flux of 33.7 Jy km~s$^{-1}$ in 2004 Dec as observed at Medicina, the 
maser flux decreased by more than a factor of $10^{3}$ in 2006 Aug, the
first GBT epoch.
By the last GBT epoch in 2007 Jan, the maser brightness had increased, but still only to
a factor of 120 less than the flux in 2004 Dec.
Continued monitoring at Medicina at 7 different epochs during 2007 and in 2008 Apr
yielded upper limits of 1--2 Jy,
confirming that the maser has remained faint since 2006.

The GBT observations with 33$''$ beam
show the maser is very close to IC1396 A:$\gamma$.
A cross pattern of observations through the source shows the offset of 
the centroid of 22 GHz maser emission from the infrared source is $11''\pm 11''$.
High-angular-resolution VLA observations by \citet{valdettaro} 
showed that the H$_{2}$O maser is actually $7''$ N of IC1396A:$\gamma$. The maser
spot is coincident with another, fainter mid-infrared source. The VLA observations were taken in 2005 Feb, when the maser was very bright (73 Jy), and the beam 
size was $0.4''$ so the shift of $7''$ is highly significant.

\clearpage

\clearpage

\subsection{2MASS\label{extinctsec}}

The 2 Micron All-Sky Survey \citep[2MASS;][]{skrutskie} Extended Mission
\footnote{http://www.ipac.caltech.edu/2mass/releases/allsky/doc/explsup.html}
 provides additional
products to the 2MASS survey. For a few regions of the sky, including
IC 1396A, 6 time exposures have been obtained. 
As a longer exposure time extends the completeness level by about 1 mag,
there is a drawback regarding the photometric saturation which occurs for
fainter sources. Consequently, photometry from the main 2MASS survey must
be preferred for all saturated sources in the 2MASS 6X catalog (i.e. when
the flag {\tt rd\_flg} equals 1).

We built the extinction map of IC 1396A following the method described in
\citet{cambresy02} using the 2MASS $H-K_s$ color excess of 32000
stars in a $40' \times 40'$ field around the nebula. Sources detected at 24
$\mu$m have been removed from the catalog since they are intrinsically red
objects for which the color is not related to the cloud extinction.
The extinction is obtained following the expression:
\begin{equation}
A_V = \left( \frac{A_H}{A_V} - \frac{A_{K_s}}{A_V} \right)^{-1} \times
(H-K_s) + {\cal Z}_{\rm col}
\end{equation}
where $(A_H/A_V - A_{K_s}/A_V)^{-1} = 15.87$ (Rieke \& Lebofsky 1985)
and $\mathcal{Z}_{\rm col}=-2.5$ mag. The zero point for the extinction
calibration, $\mathcal{Z}_{\rm col}$, corresponds to an intrinsic star
color of $H-K_s=0.16$ mag which is measured outside the nebula.
Any diffuse extinction in the field is therefore removed by construction.
The extinction is obtained using the median of the 10 stars included in each
cell. It produces a map of IC 1396A with a resolution of about 1 arcmin, 
with an average visual extinction of about 5 mag.
The extinction toward some specific Class II objects, which will be used later
in this paper are $A_{V}=5.4$, 3.8, 3.5, 4.5, and 3.5, for
IC 1396A:$\theta$, $\beta$, Tr37 11-2037, Tr37 11-2146, and 21364964+5722270,
respectively.

The extinction map clearly shows the central hole associated with
LkH$\alpha$ 349 in the higher density round head of the globule. The
visual extinction in the hole is 2.8 mag whereas it is $\sim 6$ mag in
the north side of the head and $\sim 10$ mag for the south side with a
maximum of 16.3 mag at the position 21h36m59s +57d30m00s (J2000).
Beside, the correlation of the dust extinction and the emission in the
Spitzer images exhibits a region with a deficit of infrared emission that
corresponds to the $\sim 5'$ size structure at 21h36m +57d23m, LDN 1099. This
deficit can be explained assuming LDN 1099 is disconnected from IC 1396A and
located at a larger distance of the O6 heating star. This is confirmed by
the examination of the optical images which suggests IC 1396A is illuminated
from behind whereas L1099 would be illuminated on its front size. Assuming
the system composed by the O6 star, at 750 pc, and IC 1396A is seen with an
angle of 45 degrees (0 degree meaning the star and the globule are at
the same distance), we find the main globule IC 1396A is 4 pc closer to us
than the ionizing star and that L1099 is 9 pc farther. This is confirmed by
the difference of radial velocities for the two regions which are
-8 km s$^{-1}$ and 0 km s$^{-1}$ for IC 1396A and L1099, respectively \citep{weikard}.

\subsection{Palomar Hale Telescope: Near-infrared images}

IC 1396A was imaged using the Wide-Field Infrared Camera 
\citep[WIRC;][]{wircref} on the $200''$ Hale telescope
on 2004 Aug 25-26. The 8.7$'$ field-of-view chip was 
steered toward a series of 22 positions alternating between 
a gradual scan across the globule and a nearby empty reference 
field. The reference field images were combined to generate a
sky flat, which was then removed from each image.
While observing, a small dither offset was added to each position to ensure that stars in the 11 reference field observations were
removed during creation of the sky flat by robust averaging.
The astrometry for each image was measured by comparing to
the 2MASS point source catalog. Then images were combined into
mosaics for each filter. 

The absolute calibration of the
mosaics was measured by correlating photometry in image units
to the corresponding entries in the 2MASS catalog. 
The cross-calibration is accurate to 5\%. 
The location of each of the sources discussed in this paper was 
inspected in each of the near-infrared images. 
Some sources that were marginally detected by 2MASS 
(e.g. $\alpha$, $\gamma$, $\eta$, in J band)
or suffered from blending with nearby sources
($\beta$)
were clearly detected in the Palomar images (0.9$''$ FWHM seeing).
Table~\ref{magtab} (columns J and K) 
lists the magnitudes determined from the
Palomar observations.

\begin{figure}
\epsscale{1}
\plotone{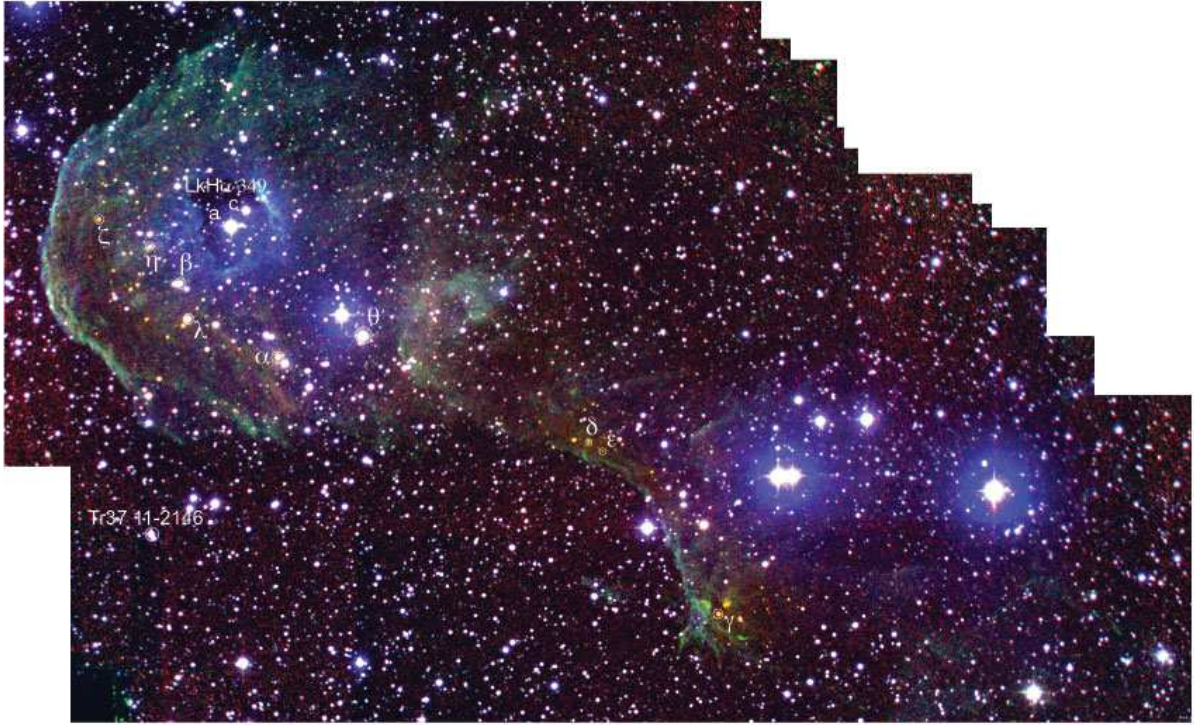}
\epsscale{1}
\figcaption[]{Near-infrared image of IC 1396A from Palomar, combining the 2.2 $\mu$m narrow continuum (red),
H$_{2}$ (green), and J-band (blue). Since the H$_{2}$ and 2.2 $\mu$m continuum wavelengths are similar,
point sources with spectral energy distribution rising to longer wavelengths appear red+green=yellow.
The Class I protostars for which {\it Spitzer} spectra are presented in this paper are labeled,
as are the LkH$\alpha$ 349 stars and the Class II star Tr 37 11-2146.
The mid-infrared sources are generally `yellow' in this representation, but they do not stand out
based on near-infrared colors alone. The vast majority of near-infrared `yellow' sources,
which preferentially fall within the globule,
are faint or not detected at 8 $\mu$m with IRAC (too faint for
mid-infrared spectroscopy with IRS); they are probably extincted background stars.
\label{ic1396apalomar}}
\end{figure}

Figure~\ref{ic1396apalomar} shows the near-infrared image. All of the bright Class I sources for which 
we obtained mid-infrared spectra are detected in this image.
Two exceptionally red sources, $\alpha$ and $\gamma$, may be
marginally resolved in the Palomar J-band image, with sizes $\sim 1.5''$ (1100 AU).
In all the Palomar near-infrared images, the protostars are dominated by
continuum, since their fluxes in line (Pa$\beta$ and H$_{2}$ 2.12 $\mu$m) and 
continuum (narrow 2.17 $\mu$m and broad J) filters
are similar.
The resolved J-band emission is likely to be scattered light.

\clearpage

\section{Class I Sources}

\subsection{Properties of the mid-infrared spectra}

Figure~\ref{classispec} shows the spectra of the bright Class I protostars
$\alpha$, $\gamma$, $\delta$, $\epsilon$, $\eta$, and $\lambda$.
Table~\ref{protoproptab} lists several properties derived from the mid-infrared
spectra including colors and silicate feature amplitude.
For the Class I sources, the silicate features at 10 and 18 $\mu$m are in absorption
hence $\Delta F/F$ is negative; conversely for most Class II sources the silicate
features are in emission.

\begin{figure}
\epsscale{1}
\plotone{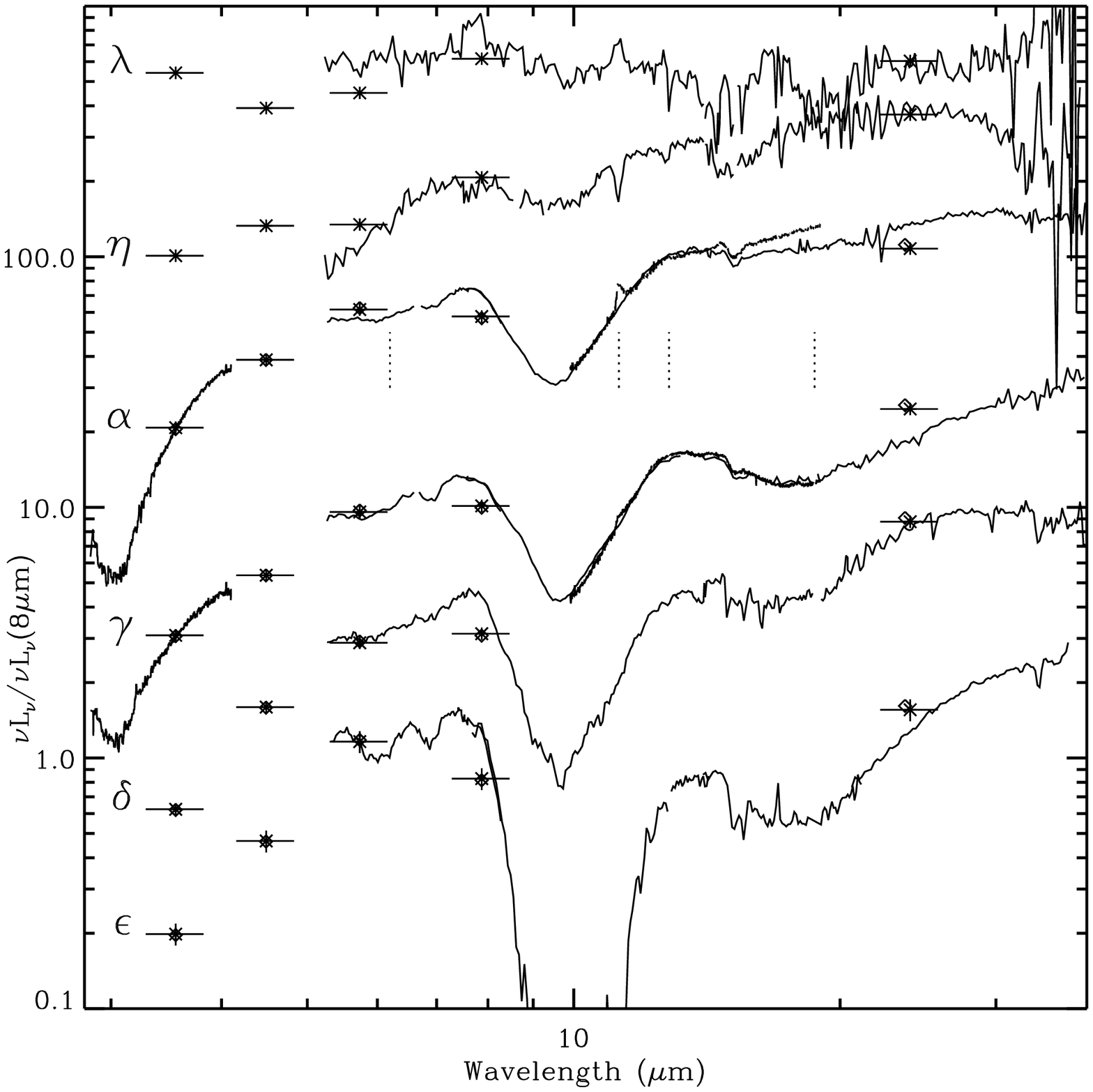}
\epsscale{1}
\end{figure}

\clearpage
\figcaption[]{{\it Spitzer}/IRS spectra of Class I sources
IC1396A:$\lambda$, $\eta$, $\alpha$, $\gamma$, $\delta$, and $\epsilon$.
Each spectrum is in $\nu L_{\nu}$ units, divided by its value
at 8 $\mu$m and offset vertically by a factor of 3.7 for clarity.
Spitzer and Palomar photometry are over-plotted as asterisks.
For sources $\alpha$ and $\gamma$, IRS high-resolution spectra and
Keck L-band spectra are included.
For source $\alpha$, the IRS high-resolution spectrum continuum
does not agree with that in the low resolution-spectrum from
15--20 $\mu$m; 
the low-resolution continuum is believed to be more accurate (and agrees
with the MIPS 24 $\mu$m photometry).
Narrow spikes in the spectra are due to rogue pixels and should be ignored.
The positive feature in $\lambda$ and $\alpha$ and negative dip in $\eta$ at 11.2 $\mu$m
are due to incomplete subtraction of a bright, nebular PAH feature; the wavelengths
of nebular lines that are incompletely removed in some spectra are indicated by dotted
lines at 6.2 (PAH), 11.2 (PAH), 12.8 ([\ion{Ne}{2}]), and 18.7 ([\ion{S}{3}]) $\mu$m.
\label{classispec}}
\clearpage

\subsection{Ice absorption features}

\begin{figure}
\plotone{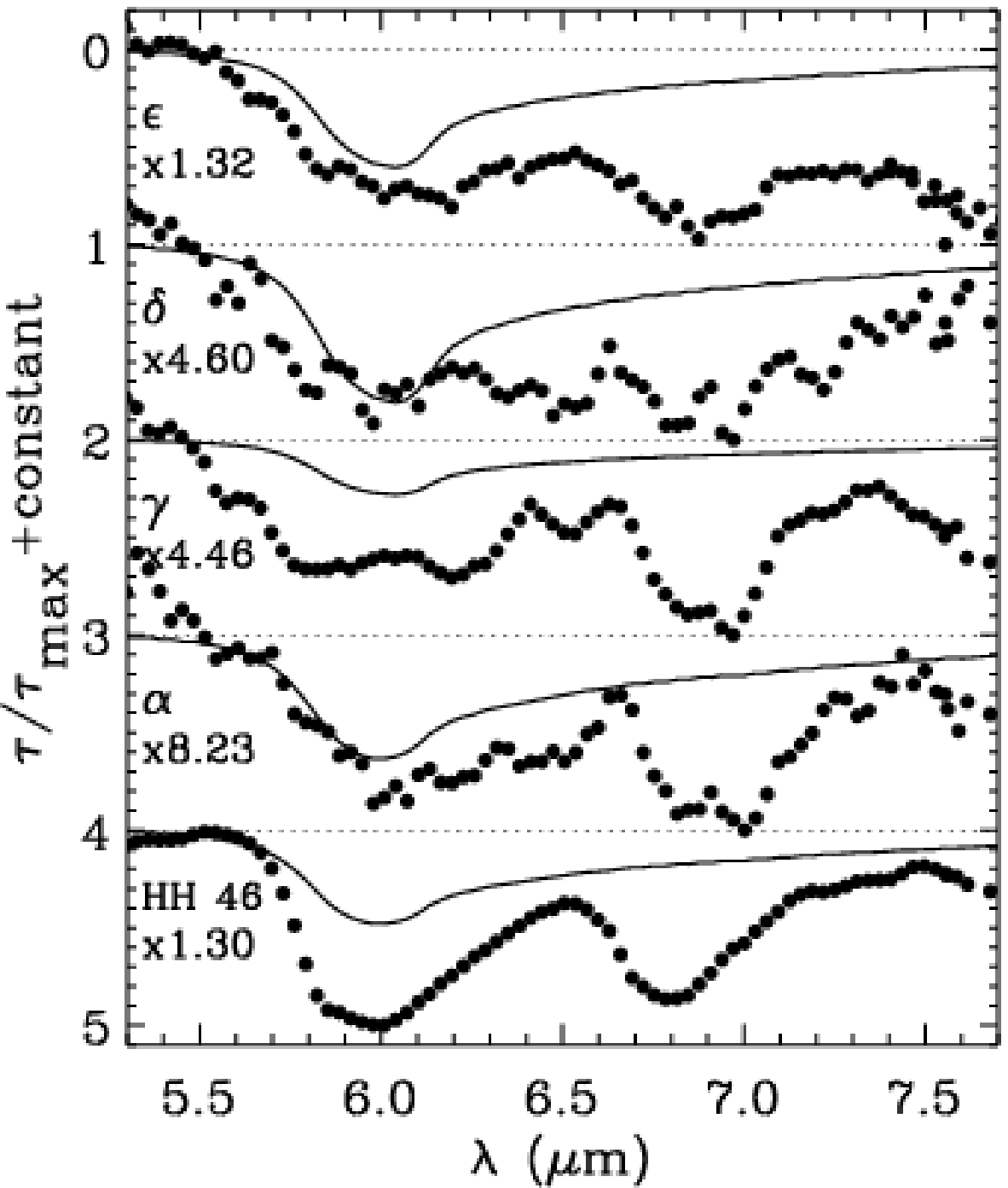}
\caption{Optical depth spectra (dots) of the Class I objects in IC 1396A showing
  5--8 $\mu$m absorption bands. The previously studied
  spectrum of the low mass YSO HH 46 IRS, representative
  of the large sample of low mass YSOs studied in \citet{boogert},
  is included at the bottom.  The
  spectra are normalized to the peak optical depth by multiplication
  with the factor indicated on the left, and then shifted along the
  vertical axis for clarity.  For each source the contribution of the
  bending mode of pure solid H$_2$O is shown by a laboratory spectrum
  (smooth line; \citealt{hud93}). Note that in all IC 1396A YSOs the
  distinct 6.85 $\mu$m\ band is shifted to longer wavelengths compared
  to HH 46 IRS.\label{f:iceplot1}}
\end{figure}

The Spitzer spectra of IC 1396A $\alpha$, $\gamma$, $\delta$, and
$\epsilon$ show the well-known ice absorption features between 5 and 8
$\mu$m, and at 15.2 $\mu$m (Fig.~\ref{classispec}). 
The latter is due to solid CO$_2$,
while the 5--8 $\mu$m absorption is caused by the overlapping bands of
many solid state species \citep{boogert}. At the long-wavelength side,
the 5--8 $\mu$m absorption overlaps with the edge of the strong 9.7
$\mu$m band of silicates, which complicates the determination of an
overall continuum level for the study of the ice absorption features.
A best-effort global continuum was determined, using all spectra and
photometry available between 1 and 40 $\mu$m, similar to the method
described in \citet{boogert}. It is kept in mind that the optical depth
of the broad absorption overlying the entire 5-8 $\mu$m range shown in
Fig.~\ref{f:iceplot1} is relatively uncertain, but that of individual,
narrower, features within it is not.

The bending mode of solid H$_2$O peaks at 6.0 $\mu$m, but is a
prominent contributor to the entire 5--8 $\mu$m absorption region. In
order to determine its contribution, the H$_2$O column density is
obtained independently from the 3.0 $\mu$m H$_2$O stretching mode,
when available (sources $\alpha$ and $\gamma$) or from the 13 $\mu$m
libration band (Table~\ref{t:colden}).  Clearly a significant fraction
of the 5--8 $\mu$m absorption is not due to the bending mode of solid
H$_2$O (Fig.~\ref{f:iceplot1}).  The excess absorption between 5.7 and
6.6 $\mu$m is likely real, and is caused by the overlapping bands of
solid H$_2$CO (5.7 $\mu$m), HCOOH (5.8 $\mu$m), NH$_3$ (6.12 $\mu$m),
and possibly ions (e.g. HCOO$^-$) and PAH species near 6.3 $\mu$m.
 
At longer wavelengths, a distinct band centered at 6.9 $\mu$m is
detected toward all sources. This ``6.85 $\mu$m band'' is commonly
observed toward low- and high-mass YSOs, as well as background
stars. Its profile is known to vary considerably and has been
decomposed in short and long-wavelengths components (called C3 and C4
in \citealt{kea01, boogert}) centered on 6.755 and 6.943 $\mu$m,
respectively.  The 6.85 $\mu$m bands toward the IC 1396A protostars are
shifted remarkably far to longer wavelengths, i.e. C4 is much stronger
than C3.  Previously \citep{boogert} it was found that such sources tend
to have low solid H$_2$O abundances, expressed in the H$_2$O column
density scaled to the peak optical depth of the observed 9.7 $\mu$m
band of silicates. 

\begin{figure}
\epsscale{.7}
\plotone{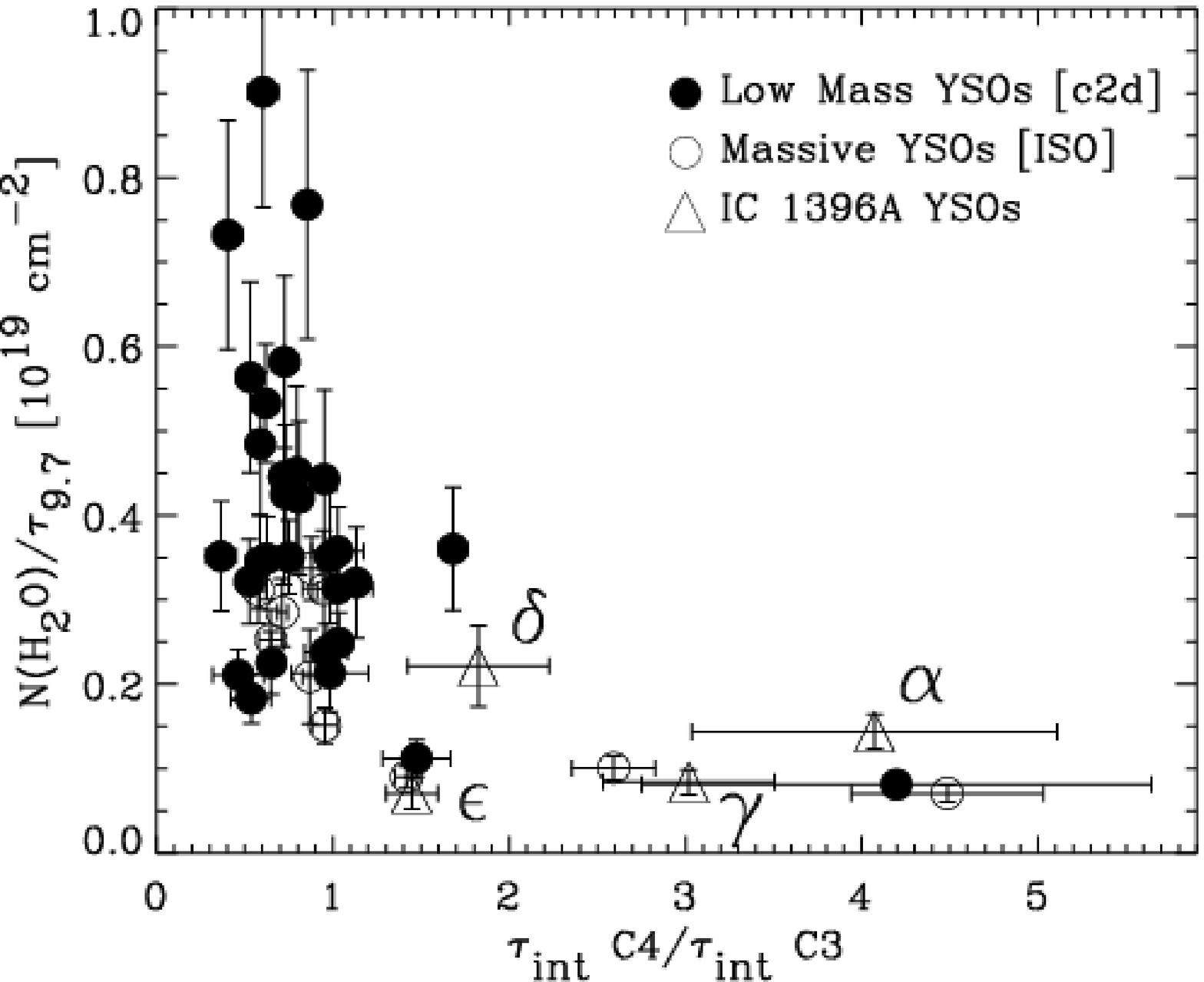}
\epsscale{1}
\caption{Correlation of the ratio of integrated optical depths of the
components C4 and C3 of the 6.85 $\mu$m bands with a measure of the
solid H$_2$O abundance for YSOs studied in the c2d ice survey 
\citep[filled circles;][]{boogert}, ISO-observed massive
protostars \citep[open circles;][]{kea01} 
and our results for IC 1396A (triangles). All 4 YSOs in IC 1396A
have strongly red-shifted 6.85 $\mu$m bands, i.e.  large C4/C3
ratios and very low solid H$_2$O abundances. This likely implies a
high degree of thermal processing of the ices in the IC 1396A core
by the nearby O-star. 
The sources with properties overlapping with the IC 1396A protostars
(C4/C3>1.2 and H$_2$O/$\tau_{9.7}$<0.25) from previous studies are 
(from high to low C4/C3): 
Mon R2 IRS3, IRAS 03301+3057, S140 IRS1, DG Tau B, and W3 IRS5.
\label{f:iceplot2}}
\end{figure}

Indeed, the IC 1396A protostars have very low $N({\rm
H_2O})$/$\tau_{9.7}$ ratios as well (Fig.~\ref{f:iceplot2}).  The
origin of the components of the 6.85 $\mu$m band is not fully
established, but the correlation with dust temperature tracers, such
as shown in Fig.~\ref{f:iceplot2}, suggests its carrier(s) are
affected by ice processing. The NH$_4^+$ ion has been considered a
reasonable candidate, as upon heating its band shifts to longer
wavelengths and its band strength increases \citep{sch03}. It is also
more refractory than H$_2$O ice.  Regardless of its origin, it appears
that the ices toward the IC1396A protostars are thermally, and possibly
energetically processed. All 4 studied YSOs show these unusual ice
characteristics, which is distinct from most sources in the large
sample of low mass YSOs in nearby star-forming regions 
studied by \citet{boogert}. This suggests that
the ices toward IC 1396A are processed by radiation external to the
YSOs, likely the harsh radiation from the nearby O star.

Finally the solid CO$_2$ abundances relative to H$_2$O are between
14-34\% (Table~\ref{t:colden}), which is typical for most YSOs
\citep{pon08}. Solid CH$_3$OH was looked for at 9.7 and 3.53 $\mu$m as
well, but is less than 5\% with respect to H$_2$O.

\def\oldversion{
Absorption features due to CO$_{2}$ (15.2 $\mu$m) and H$_{2}$O (6 $\mu$m), 
as well as the unidentified but
frequently-observed feature at 6.9 $\mu$m, are evident in the Class I sources.
Figure~\ref{iceplot} shows the absorption spectra from 5.2--8.1 $\mu$m containing the 6 $\mu$m H$_2$O feature,
and Figure~\ref{iceplot15} shows the absorption spectra centered on containing the 15.2 $\mu$m CO$_2$ feature.
Table~\ref{protoproptab} lists the ice absorption features strengths.
The ice absorption features are correlated, with the average depth of CO$_{2}$ divided by depth of
the 9--11 $\mu$m silicate feature being 0.33. The H$_{2}$O feature is not detected in the weakest
sources; a linear correlation with the silicate depth yields a slope of 0.33 (the same as for
CO$_{2}$) but with a significant offset as if the H$_{2}$O is only present for sources with
silicate feature depth greater than 0.18. 
Since the protostars are embedded within dense parts of the globule, some of the silicate
feature (and possibly also ice) could be due to the foreground.
Since the interstellar extinction curve has silicate feature depth $\sim 0.08 A_{V}$ and the
estimated extinction $\sim 1.5$--3.5 (based on the SED fits for the Class II sources), 
we expect foreground extinction due to the globule to give rise to
silicate feature depths $\sim 0.08$--0.18 (taking into account the way we subtracted
baselines when measuring the silicate depth on the Spitzer spectra).
If we assume the globule extinction has no associated ices and remove it from the fit,
the ratio of CO$_{2}$/silicate and the ratio of H$_{2}$O/silicate absorption are both 0.33.

Besides the 6 $\mu$m and 15 $\mu$m features, the 6.9 $\mu$m feature is detected with depth
comparable to and proportional to the 6 $\mu$m feature. 
We searched for NH$_{3}$ (9.0 $\mu$m) and CH$_{3}$OH (9.7 $\mu$m) features such as 
were seen toward W33A \citep{gibb}, but they are not present for the IC1396A protostars;
the upper limit on $\Delta F/F$ is 0.05. There is a hint of CH$_{3}$OH for $\alpha$ at a
level of 0.03. These features are at the bottom of the 9--11 $\mu$m silicate feature. For
source $\epsilon$, there is essentially zero light (to within measurement accuracy) transmitted
by silicates, so a measurement of these ice features is practically impossible.

\begin{figure}
\epsscale{1.1}
\plotone{iceplot}
\epsscale{1}
\figcaption[]{Spectra from 5--8 $\mu$m of Class I sources in IC 1396A showing
the 6 $\mu$m H$_{2}$O and 6.9 $\mu$m unidentified ice features.
The sources are plotted in order of silicate absorption depth.
\label{iceplot}}
\end{figure}

\begin{figure}
\epsscale{1.1}
\plotone{iceplot15}
\epsscale{1}
\figcaption[]{Spectra of Class I sources in IC 1396A centered on the 15.2 $\mu$m CO$_{2}$
ice feature.
\label{iceplot15}}
\end{figure}
}

\subsection{Models for the Class I protostars}

We now attempt to derive the physical properties of the Class I protostars
from the observed spectra and photometry. To this end, we first create a
simple `toy' model of a Class I object and use it to derive some
fiducial parameters for the envelope. 
The simple model is a combination of modified blackbodies, with outer layers
absorbing inner ones such as was applied to protostars in
the Trifid nebula \citep{rhotrifid}.
Then we compare the observations to 
more sophisticated radiative transfer models developed by 
\citep{whitney}, which allow us to estimate the properties of the central stars
and their accretion rate.
While the theoretical models are more physically detailed,
they don't fit the spectra as accurately as the simple models, which
we present here as an empirical description of the data.

\subsubsection{Simple model}

The spectral energy distributions of the Class I sources appear to be the sum of two primary 
color temperatures, with the warmer component having $T_{1}\simeq 420$ K and a colder 
component having $T_{2}\simeq 120$ K. To explain the 70 $\mu$m emission,
 a colder component is required,  with $T_{3}\simeq 40$K.
As a first, empirical, description of the 
spectra, we will model each protostar as the sum of three nested components, with
the higher temperature ones residing closer to the center and being absorbed
by the colder ones:
  \begin{equation}
 F_{\nu}^{obs} = F_{\nu}^{1} e^{-(\tau_{2}+\tau_{3})} + F_{\nu}^{2} e^{-\tau3} + F_{\nu}^{3}.
 \end{equation}
This simple model neglects direct emission from a potential central star, as well as neglecting
details of radiative transfer. Neglecting the central star is entirely justified based
on the lack of optical counterpart and the great optical depth toward the center.
Neglecting radiative transfer is not justified, so one must bear in mind that the empirical
models here are only descriptions of the data and not necessarily self-consistent.
(More sophisticated models are described in the next section.)
The optical depth of the outer layers was normalized to match the observed 
depth of the 9--11 $\mu$m silicate feature, and the wavelength-dependence of the opacity
was taken from the interstellar dust model of \citet{draineLi}.
The coldest component (3) was fixed at 40 K and
was normalized to the difference between the observed 70 $\mu$m flux and the model
for components 1 and 2 (the inner and outer envelope, respectively).

\begin{figure}
\plotone{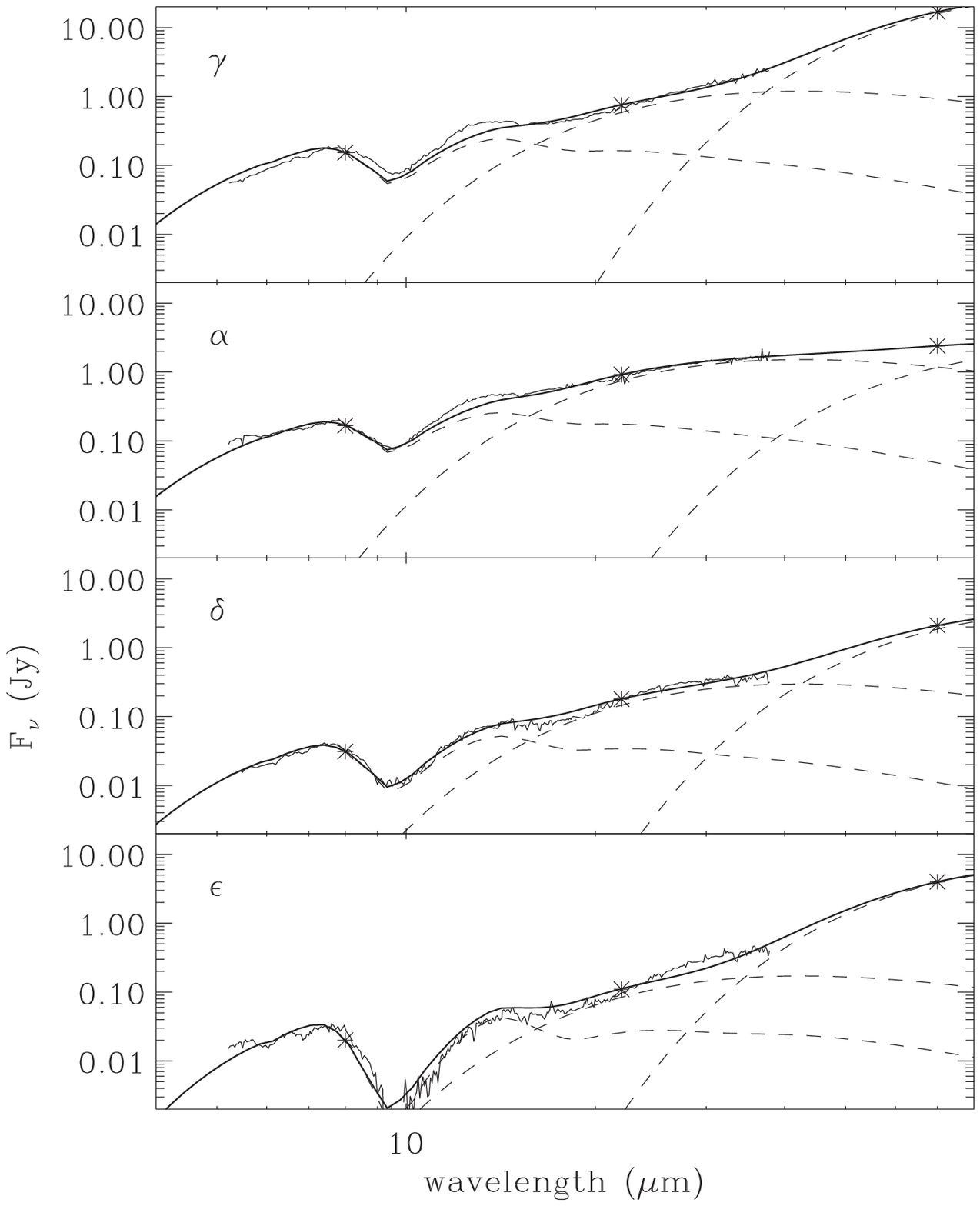}
\figcaption{Empirical model fits to the Class I protostar spectra. For each source a sum
(thick solid line)
of three blackbodies (shown individually as dashed lines), with the warmest one absorbed by
the cooler ones, matches the observed IRS spectra (thin solid line) and
MIPS photometry (asterisk).
\label{protomodel}}
\end{figure}

Figure~\ref{protomodel} shows the observed spectra and empirical model fits. 
The color temperatures and ratio of the components 1 and 2 were adjusted to 
match the spectrum from 5--40 $\mu$m. 
We found it possible to match all the spectra with similar values of $T_{1}$ and $T_{2}$,
so we fixed them at 420 and 120 K, respectively. Specifying the 22 $\mu$m flux,
$F_{22}$, the ratio of fluxes at 22 to 8 $\mu$m, $F_{22}/F_{8}$, and the silicate
feature optical depth, $\tau_{sil}$ was sufficient to constrain components 1 and 2.
Greybody dust grains in region (2) heated by radiation 
from region (1) with luminosity $L_{1}$ will be at temperature 
\begin{equation}
T_{2}=278 \left(\frac{L_{1}}{L_{\odot}}\right)^{1/4} \left(\frac{R_{2}}{{\rm A.U.}}\right)^{-1/2}
\,\,{\rm K}.
\end{equation}
Combining the outer envelope (region 2) optical depth and size, and assuming dust/gas abundance as in the ISM,
we then derive the density, $n_{2}$, in region 2, which is of order $10^{9}$ cm$^{-3}$.
If the protostar is in steady accretion, then its density should follow
\begin{equation}
n_{2} = \frac{\dot{M} R_{2}^{-3/2}} {4 \pi \mu m_{H} \sqrt{2 G M_{*}}},
\end{equation}
\citep[e.g.][]{terebey}
from which we derive mass accretion rates
of order $10^{-6} M_{*}^{1/2}$ $M_{\odot}$~yr$^{-1}$.
The accretion luminosity of the outer envelope alone (i.e. for initially distant material that
only moves $R_{2}$) is only of order 0.1 $M_* L_{\odot}$, less than
the observed luminosity of the outer core ($L_{2}\sim 1 L_{\odot}$). 
This shows that material must be accreting further in than $R_{2}$.
If the accretion luminosity arises at a radius $R_{acc}$, and
there is a steady flow of matter inward, then
\begin{equation}
L_{core} = \frac{G M_{*} \dot{M}_{env}}{R_{acc}},
\end{equation}
which yields accretion radii $R_{acc}\sim 30 M_{*}$ AU. 

The coldest component (3), or outer envelope, was fixed at 40 K and
was normalized to the difference between the observed 70 $\mu$m flux and the model
for components 1 and 2. including for relevant component $i$ the visible extinction, $A_{V}^{i}$, the
distance from the center, $R_{i}$, the volume density, $n_{i}$, and the luminosity, $L_{i}$.
The opacity of the outer envelope was small, so the silicate absorption can be
identified with component 2, the envelope. The volume density of the outer envelope is only an order
of magnitude higher than that of a dense core \citep{myersref}, so it represents
the outermost layer of the protostar that merges with the surrounding globule gas.

Table~\ref{protomodeltab} lists the results of the model fits,
In this empirical model, a collapsed object of
radius $<30 R_{\odot}$ has formed, and it resides inside a spherical,
optically thick region of size $\sim 2$ AU.
The mass of the star cannot be readily constrained without further guidance
from theoretical models. We assume masses of order 1 $M_{\odot}$ for the 
calculations in this section. Changing to larger or smaller masses has
no effect on the luminosities (which are tied to the observed fluxes), 
but mostly changes the size of the central object (which is hidden within
the optically thick core). 

The luminosity of sources $\alpha$ and $\delta$ arises primarily from components (1) and (2), 
which is consistent with
these objects harboring a central protostar in the Class I phase.
The luminosity of component (3), the outer envelope, is highest for $\gamma$ and $\epsilon$.
The key measurement is the 70 $\mu$m flux: the ratio of 70/24 $\mu$m flux is
22, 2.6, 12, 36 for $\gamma$, $\alpha$, $\delta$, and $\epsilon$, respectively.
The 70 $\mu$m fluxes are uncertain due to the bright nebular emission and large beam;
however they are sufficiently accurate to reveal differences between the sources.
(Sources $\delta$ and $\epsilon$ are not cleanly resolved from each
other at 70 $\mu$m; the peak pixel brightnesses are comparable [with $\epsilon$
being $\sim 7$\% brighter], 
so we attributed half the 70 $\mu$m flux to each source.)
The origin of the 70 $\mu$m flux from $\gamma$ and $\epsilon$ may be different.
For $\gamma$, the presence of an H$_2$O maser spot $7''$ from the infrared source
(see \S\ref{gammassec}) 
indicates a Class 0 core may contribute some of the 70 $\mu$m flux we attributed to $\gamma$.
For $\epsilon$, the high 70/24 $\mu$m flux (and correspondingly, the high luminosity of component (3)
in the simple model) is due to the source being more envelope-dominated, since the
silicate feature is extremely deep for this source.

\def\old{
For IC 1396A:$\gamma$, $\tau_{sil}\simeq 1.7$ and the envelope extinction $A_{V}^{env}=24$.
The luminosity of the core, before being absorbed by the envelope, $L_{core}\sim 3.0 L_{\odot}$.
Greybody dust grains in the envelope at temperature $T_{env}$ heated by $L_{core}$ would be 
at a distance $R_{env}\sim 3.1$ AU from the core.
Combining the envelope optical depth and size, its volume
density $n_{env}\sim 1.0\times 10^{9}$ cm$^{-3}$.
If the envelope is in steady accretion, then its density should follow
\begin{equation}
n_{env} = \frac{\dot{M}_{env} R_{env}^{-3/2}} {4 \pi \mu m_{H} \sqrt{2 G M_{*}}},
\end{equation}
\citep[e.g.][]{terebey}
which requires an accretion rate
$\dot{M}_{env}\sim 2.3\times 10^{-6} M_{*}^{1/2}$ $M_{\odot}$~yr$^{-1}$.
The accretion luminosity of the envelope alone (i.e. for initially distant material that
only moves to the envelope radius) is only 0.1 $M_* L_{\odot}$, much less than
the observed luminosity of the envelope (2.9 $L_{\odot}$). 
This shows that the envelope must be heated from within.
If we require the accretion luminosity to arise in the core, {\it and}
there is a steady flow of matter from the envelope, then
\begin{equation}
L_{core} = \frac{G M_{*} \dot{M}_{env}}{R_{acc}},
\end{equation}
which, given the values already derived, requires the accretion radius 
$R_{acc}\sim 10 M_{*}$ AU. Thus in this empirical model, a collapsed object of
radius 10 $R_{\odot}$ must have formed and resides inside a spherical,
optically thick region of size $\sim 2$ AU.
}

\begin{figure}
\epsscale{.5}
\plotone{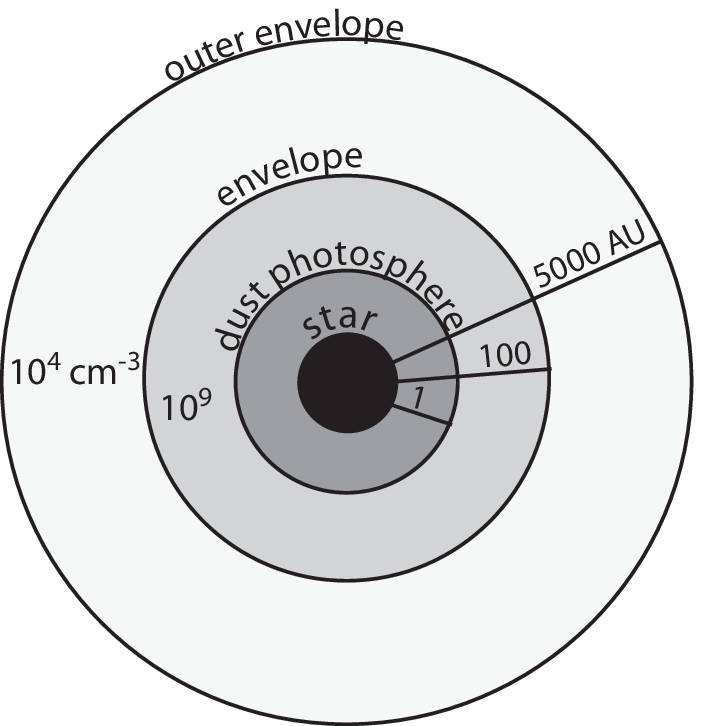}
\epsscale{1}
\figcaption{Cartoon illustrating the structure of the Class I protostars. The mid-infrared
emission detected by {\it Spitzer} is attributed to the `dust photosphere' spanning 
1--100 AU and surrounded by a cooler envelope.\label{protocartoon}}
\end{figure}

We can now compare this empirical, two-component description of the spectral
energy distribution to a simple theoretical model for the protostar. Following
\citet[][hereafter SP]{spbook}, we expect a dust photosphere, located where the protostar is optically
thick to its emergent radiation. This `photosphere' will span a range of radii due to the rapid
wavelength dependence of the dust opacity. To determine whether the mid-infrared 
emission can be associated with such a photosphere, we 
solve SP equations 11.12a,b using the luminosity 6 $L_{\odot}$ (like source $\alpha$) and
an assumed mass 1 $M_{\odot}$. The photospheric temperature is in the range 
500 -- 120 K for a mass accretion rate in the range
0.15  -- 4$\times 10^{-5} M_{\odot}$, and
a photospheric radius in the range 1.4-120 AU, respectively.
These temperatures span the `core' and `envelope' color temperatures required to
explain the 5--40 $\mu$m emission. 
We furthermore note that the `envelope', to which
we ascribed the 9--11 $\mu$m silicate absorption, has, by definition, an optical depth
greater than 1 (but not greater than 10) at 10 $\mu$m, so by the simplest definitions it
must be considered part of the photosphere. That we required two color temperatures
to match the spectral energy distribution, rather than a single blackbody, reflects only
that the wavelength-dependence of the opacity leads to an extended photosphere.
By ascribing the 9--11 $\mu$m silicate absorption to the cooler component, we have
implicitly assumed a negative temperature gradient, as anticipated for a source powered
by accretion luminosity rather than being heated by starlight and cosmic
rays from the outside.
Figure~\ref{protocartoon} summarizes the protostar parameters derived in this section.
The agreement of the accretion rate for a simple photosphere approximation with that
derived in the previous paragraph from the envelope density,
and the plausibility of the protostellar radius required for accretion to provide
the observed luminosity,
suggest the simple protostar model presented here is at least roughly consistent with
growing star surrounded by a thick envelope and powered by accretion.

\subsubsection{Association with Whitney protostar models\label{classiwhitney}}

Detailed radiative transfer models of protostars with realistic geometries 
have been developed by \citet{whitney} and 
calculated over a wide range of physical parameters by \citet{robitaille}.
To find the Whitney model that best matches the Class I source spectra, we
first tabulated, for the observed spectra for sources $\alpha$, 
$\gamma$, $\delta$, $\epsilon$, $\zeta$, and $\eta$ at a set of 14 wavelengths
(1.22, 1.63, 2.19, 3.56, 4.51, 5.58, 7.65, 9.95, 12.93, 17.72, 24.28, 
29.95, 35.06, 71.42 $\mu$m).
These wavelengths were chosen to sample the SED and largely avoid ice features;
however, the 12.93 $\mu$m feature possibly contains a strong H$_{2}$O libration feature
which is not included in the models \citep{boogert}.
For each of the $2\times 10^{5}$ models from \citet{robitaille},
we used the model fluxes integrated over the  J, H, K, IRAC 3.6,  IRAC 4.5, and MIPS 
70µm filters\footnote{available at http://caravan.astro.wisc.edu/protostars/repository/index.php}, and individual 
model fluxes at the remaining Spitzer IRS wavelengths provided separately by T.~Robitaille (personal communication).
A $\chi^2$ goodness-of-fit was calculated from the
difference between the observed and model tabulated fluxes, including for each protostar model
10 possible inclination angles (87.1$^\circ$ [edge-on], 81.4, 75.5, 69.5 ,63.3, 56.6, 49.5, 41.4, 31.8, 18.2 [pole-on]),
variable ($1<A_V<8$ mag) line-of-sight reddening applied with the Milky Way $R_V=3.1$ extinction curve 
\citep{weingartner}, and a normalization factor within the range $\pm 20$\% relative to that
appropriate for a distance of 750 pc. This procedure is very similar to the one available
from Robitaille's protostar model fitter$^3$, and we have verified for source IC1396A$\alpha$ that
the model identifications are in agreement.
The goodness-of-fit ($\chi^{2}$) was often poor, potentially due to coarseness and inapplicability of the model grid, 
and possible source multiplicity (IC1396A:$\gamma$ discussed below).
In many instances, particularly for sources IC 1396A: $\delta$ and $\epsilon$, the large 
$\chi^2$ is caused by a steeply rising MIPS 70 $\mu$m flux 
that was not reproduced by the models, although the rest of the SED was fitted very well.
These sources are located in the southwestern rim of IC 1396A and have large fitted extinctions of 
$A_V\sim~7$--8 mag. 
We suspect that outer envelope heating by the external UV radiation field that is not accounted for in the Robitaille models is  the cause for the enhanced MIPS 70µm fluxes. 

\begin{figure}
\epsscale{1}
\plotone{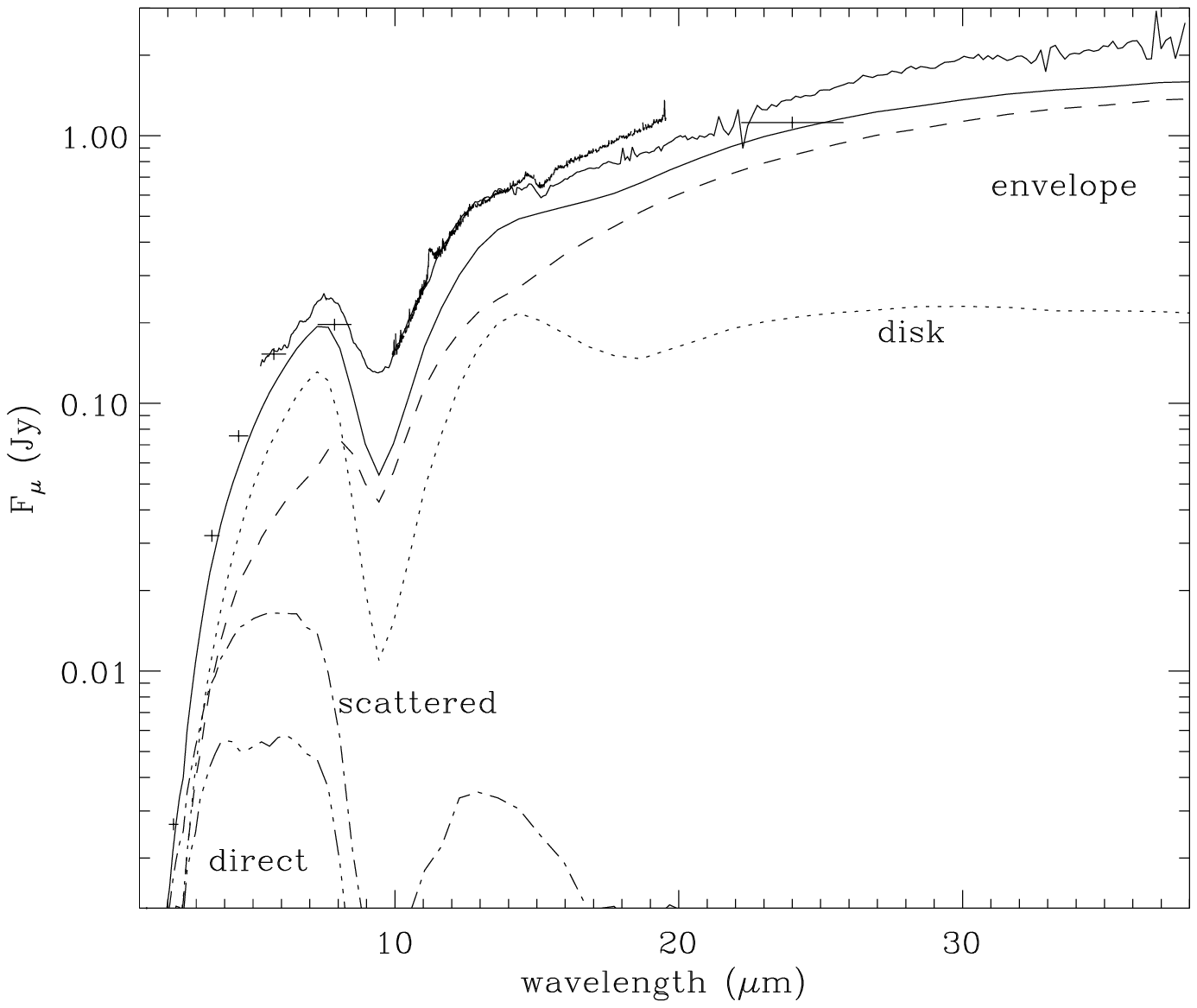}
\epsscale{1}
\figcaption[]{Spectrum of IC1396A$\alpha$ together with its best-fitting radiative
transfer model from among those computed by \citet{robitaille}. The IRAC data are symbols
with error bars; the IRS spectrum is the solid, jagged lines. Model curves are scaled
downward by 10\% for clarity.
The model sum is a solid,
smooth line. Individual model components are the outer envelope (dashed line) that contributes in
the far-infrared,
the inner envelope or `disk' (dotted line) that contributes in the mid-infrared flux,
scattered light (dash-dot line), and direct light (dash-dot-dot-dot).
\label{plotwhitney}}
\end{figure}

The 10 models with the best $\chi^{2}$ were inspected in
detail for each source. Table~\ref{whitneytab} shows the most important parameters
for the best-fitting model for each source.
Figure~\ref{plotwhitney} compares the radiative transfer model to the observed spectrum
of IC1396A:$\alpha$.
At wavelengths longer than 10 $\mu$m , the Class I protostar models are
dominated by the `envelope,' while at 5--8 $\mu$m the flux is mostly 
from the `disk' in these models. 
The 9--11 $\mu$m silicate feature is
due both to absorption of the `disk' by the envelope and radiative transfer
through the envelope (whose temperature decreases radially outward).
Scattered light contributes significantly from $\sim 2$--4 $\mu$m, and
emergent flux from the central star
becomes dominant at wavelengths less than 2 $\mu$m. 
It is immediately evident
that these radiative transfer models are more realistic and contain more of the
appropriate physics than the simple model in the previous section:
no direct flux from the star would be detectable in the simple, spherical model; nor
is the radiative transfer through the envelope taken into account.
The detection of these sources in the near-infrared demonstrates that the
envelope and/or disk must be flattened \citep[or inhomegenous; see][]{indebetouw}, 
in order to allow scattered light
from the star to emerge.

The properties of the central stars are not well constrained, but
the model fits clearly prefer the youngest available stellar photosphere
models, with ages or order $10^3$ yr for $\alpha$, $\delta$ and $\epsilon$,
$10^4$ yr for $\gamma$ and $\eta$, and $10^5$ yr for $\zeta$.
The evident progression of SED shapes in Figure~\ref{classispec} 
agrees with the model fitting results, suggesting that the SED shape 
is  sensitive to stellar age. 
The central
star only contributes significantly to the K-band and 3.6 $\mu$m IRAC
band, and it is highly extincted. Taken at face value, the masses are
similar, at 0.2--0.3 $M_{\odot}$. 
Since the accretion rates are still
high, these may not be the final masses of the young stars that will
evolve toward the main sequence. If the sources remain at this
accretion rate for a nominal Class I lifetime of $\sim 10^5$ yr,
the added mass would be in the range 0.2-0.9 $M_{\odot}$, suggesting
the final masses of the stars would be in the range 0.6-1.1 $M_{\odot}$.

\subsubsection{Comparing the empirical and radiative transfer model results}

For sources IC1396A:$\alpha$ and $\epsilon$, the total luminosities obtained
from the empirical (Tab.~\ref{protoproptab}) and radiative transfer (Tab.~\ref{protomodeltab})
are in accord. 
The discrepancy in luminosities for $\gamma$ is due to lack of an outer envelope in the radiative transfer model,
or presence of a Class 0 source contaminating the 70 $\mu$m flux of $\gamma$, as discussed above.
The discrepancy in luminosity for $\epsilon$ is due to lack of an outer heated envelope
in the radiative transfer model. The accretion of the envelope is not included as
a source of heating \citep[see \S2.2.4 of][]{robitaille}.

The accretion rates, which are probably the most interesting quantities
as they tell us how rapidly the stars are growing, are in the same order of magnitude 
($10^{{-6}} M_{\odot}$~yr$^{-1}$).
In both the empirical and radiative transfer models, 
the mid-infrared flux is dominated by the envelope, whose
luminosity is determined by the accretion rate; thus the
accretion rates are fairly well constrained by the observations.
Nonetheless, there are some significant differences.
In particular, the empirical model yields an accretion rate 9 times higher
than the best-fitting radiative transfer model for $\epsilon$.
The massive outer envelope contributes 80\% of the emergent
luminosity in the empirical model for source $\epsilon$.
The luminosities are also in general agreement, as expected since
the models were chosen to match the observations at wavelengths
that span the brightest part of the spectral energy distribution. 
For source $\gamma$, 
the empirical model yields a significantly higher luminosity, due
to the cold, outer component that was required to match the 70 $\mu$m
flux. The Whitney models cannot match the 70 $\mu$m flux, hence the 
poor fit quality for $\gamma$.  

\clearpage

\section{Class II Sources}

\subsection{Disk Properties from the mid-infrared spectra}

\def\extra{
temperature, silicate emission feature, luminosity of disk,
Ldisk/Lstar
num1		200
num6		200
Tr37		200
}

Figure~\ref{classiispec} shows the spectra of four Class II sources.
Each spectrum was dereddened using the line of sight 
extinction derived in \S\ref{extinctsec}
and a nominal $R_{V}=3.1$ extinction curve \citep{weingartner}.
The extinction correction is small in the mid-infrared:
for $A_{V}=5.4$, the optical depth $\tau=0.14$ at the shortest IRS
wavelengths (5.5 $\mu$m), 0.38 at the peak of the interstellar silicate feature
(9.5 $\mu$m) and 0.06 at long wavelength (30 $\mu$m). 
But the line of sight extinction is high enough to block much of the optical and nearly
all ultraviolet light from these young stars, making it impossible to assign  
spectral types using the stars' colors.

The mid-infrared spectra in Figure~\ref{classiispec}
are characterized by a modest near-infrared brightness, a
very flat continuum from 3-8 $\mu$m, a bright silicate emission
feature at 9--11 $\mu$m, and a continuum beyond 15 $\mu$m
with a range of brightnesses.
Table~\ref{protoproptab} lists the colors and silicate feature
strengths for the sources. 
The silicate feature ranges from very strong (twice the continuum)
to weak (20\% of continuum) to undetectable.

The radial and vertical structure of the disks can be complicated
during the time of planet-building due to ejection of
material from the planet growth zones as well as collection of material
near stable resonances. Of particular importance for
understanding formation of the Earth is the presence
or lack of solid material in the $\sim 0.5$--2 AU `habitable'
region. 
For all the Class II sources with significant emission at 24 $\mu$m,
the color temperature of the 15--40 $\mu$m emission is around 200 K.
There is no evidence of the presence of a colder disk. Given luminosities
of $L_{*}$, the color temperatures suggest that the emission
arises within 1.9 $L_{*}^{1/2}$ AU of the star, corresponding to
the zone of terrestrial planet formation.
Emission also arises from 5--10 $\mu$m,
requiring dust even closer to the star.

\begin{figure}
\epsscale{1}
\plotone{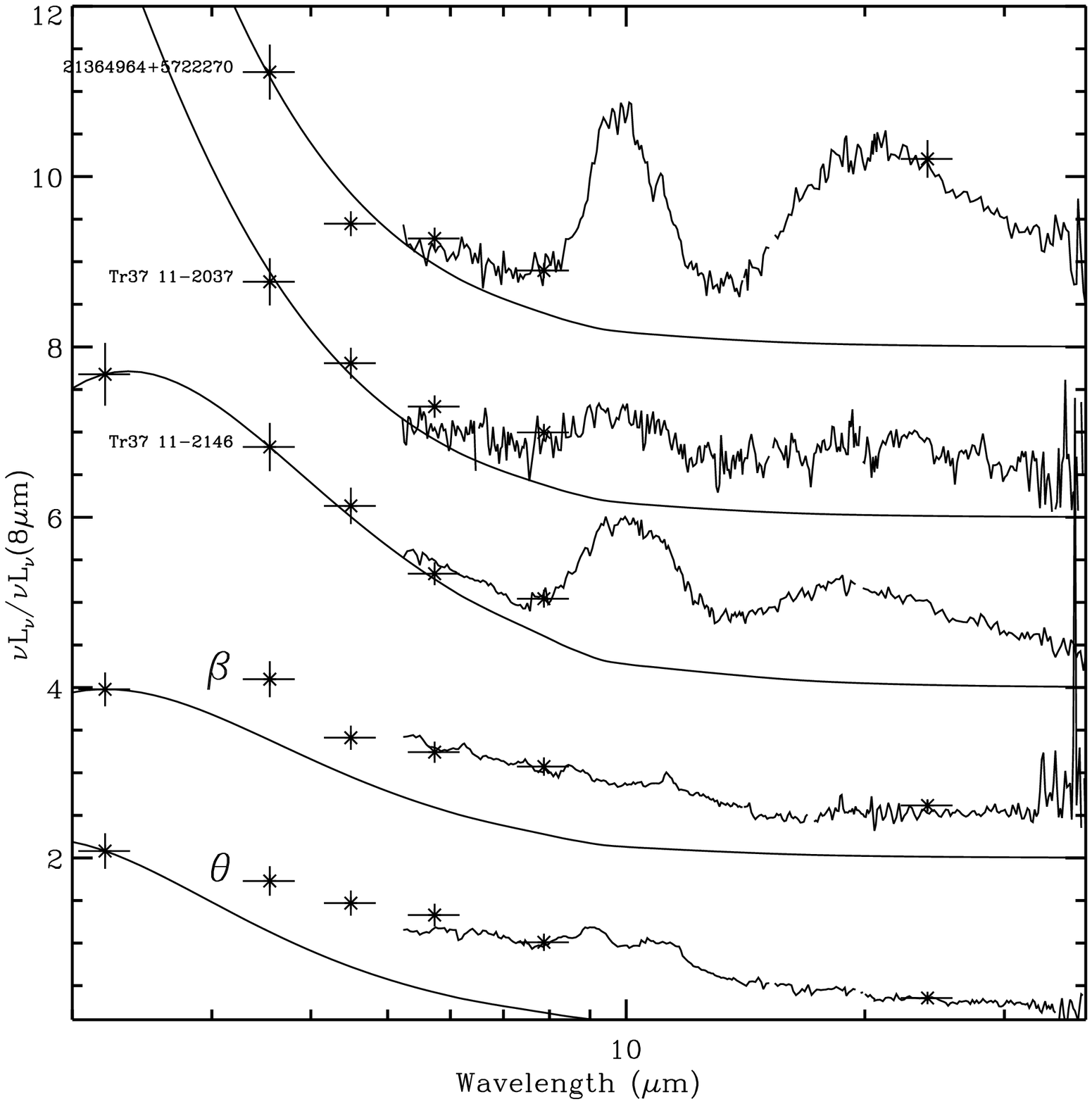}
\epsscale{1}
\end{figure}
\clearpage
\figcaption[]{{\it Spitzer}/IRS spectra of Class II sources
in IC 1396A.  Each spectrum is in $\nu L_{\nu}$ units, divided by its value
at 8 $\mu$m and offset vertically by 0, 2.5, 5, 7.5, 10 $L_{\odot}$,
for sources IC1396A:$\theta$, $\beta$, Tr37 11-2146, Tr37 11-2037, and 21364964+5722270,
respectively. The solid line is a simple blackbody through the near-infrared photometry to
illustrate the approximate flux of the photosphere.
\label{classiispec}}
\clearpage

Mid-infrared spectra of large samples of T Tauri and Herbig AeBe stars have been presented by
\citet[][IRS: Taurus-Auriga]{furlan}, 
\citet[][IRS: Chameleon, Lupus, and other nearby star-forming regions]{kessler},
\citet[][Subaru: Taurus, TW Hydra]{honda}, 
and  \citet[IRS: Tr 37 and NGC 7160]{sicilia07}.
Our spectra of Class II sources in IC 1396A, 
selected for 8 $\mu$m brightness, are generally similar to
T Tauri stars in other star-forming regions.
\def\extra{
None of our sample present the distinctive
shape of CoKu Tau 4, where the spectrum is dominated by the
star out to 8 $\mu$m, followed by a weak silicate feature
and then a rapidly-rising disk with a
relatively cold color temperature (135 K). 
That disk gaps like in CoKu Tau 4, if due to planet formation,
are not present in our sample may suggest that inner planets have not yet formed
around these stars; however, a recent observation shows CoKu Tau 4 is a binary
star \citep{ireland}, so the lack of similar stars in our sample may have no
bearing on planet formation.
}
Silicate emission features are present in emission for all 5 stars in Figure~\ref{classiispec}.
There are two evident morphologies of silicate features. The top 3 spectra show
bright, distinct 9--11 and 18--22 $\mu$m features similar to other T Tauri stars. 
These features are produced by small amorphous silicate grains in an optically-thin portion
of the disk, e.g. its upper surface or a flared or warped region illuminated by
the star.

The bottom two stars in Figure~\ref{classiispec},
IC 1396A:$\theta$ and $\beta$, have much weaker and broader
features at 9--11 $\mu$m and no evident feature at 20 $\mu$m.
Both sources are bright in the IRAC and MIPS images;
in the near-infrared (Fig.~\ref{ic1396apalomar}), both have have nearby confusing sources.
Structured nebulosity could contaminate the spectra near the PAH and important
ionic lines. 
$\theta$ may have
associated scattered light in the optical (but not in our Palomar or IRAC images). 
We applied local background subtractions when extracting the spectra,
to minimize any contamination from the nebula or neighboring sources. 
The silicate features for these stars have low amplitude, but
there are two peaks that are
evident at the long and short-wavelength edges, making the feature appear `horned.' 
The `horned' appearance is not an effect of line-of-sight extinction.
The interstellar absorption feature peaks at 9.5 $\mu$m and is broad.
If the extinction to IC 1396A:$\theta$ were $A_{V}>10$, then any silicate
emission feature from the star would have its 9--10 $\mu$m portion depressed,
generating a very ``red'' emission feature containing only the $\lambda>10$ $\mu$m
portion.
For IC1396A:$\theta$, the two `horns' are at 9.1 and 11.3 $\mu$m. The 11.3 $\mu$m horn is
at the appropriate location for olivine, Mg$_{x}$Fe$_{1-x}$SiO$_{4}$. We now consider two
possible explanations for the shape of the silicate feature.

\begin{figure}
\epsscale{0.8}
\plotone{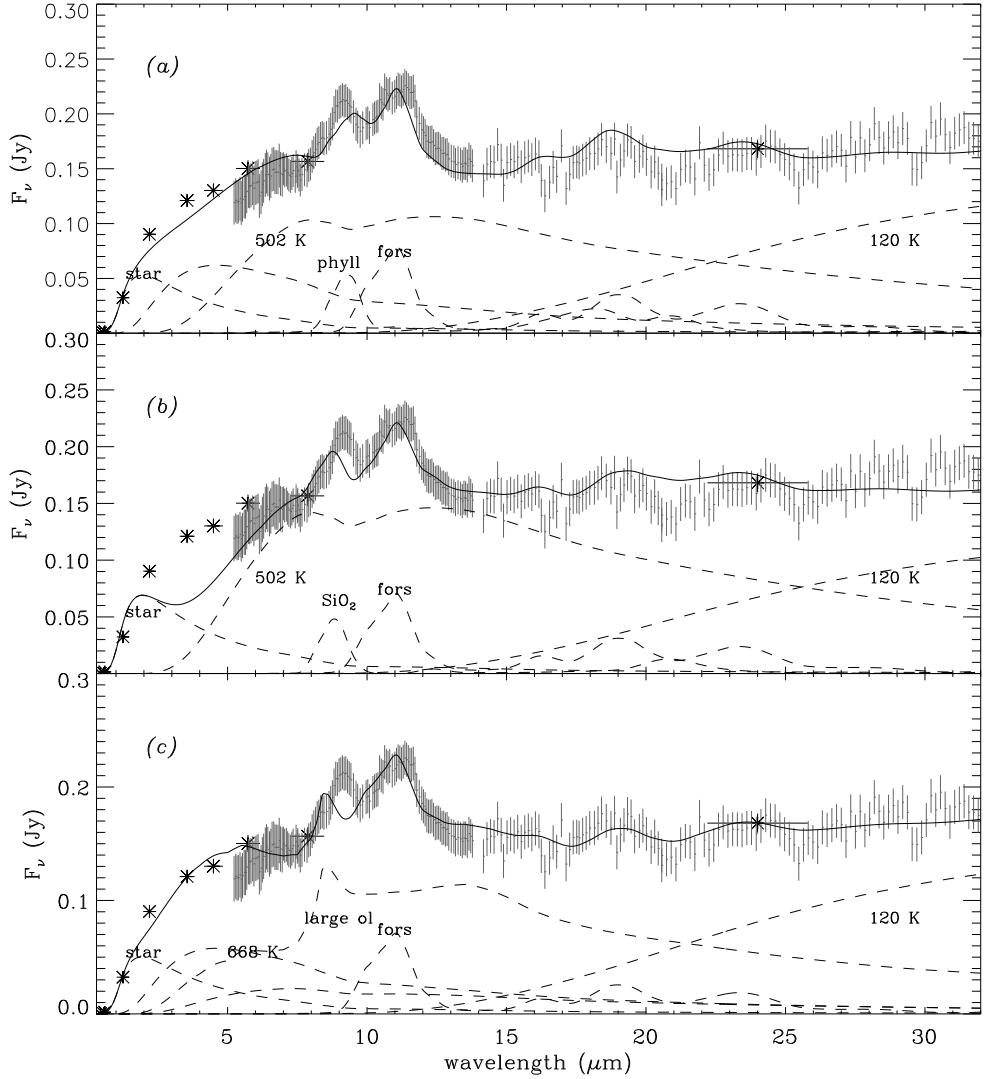}
\epsscale{1}
\figcaption[]{Spectrum of IC 1396A:$\theta$ together with three mineralogical fits.
Panel {\it (a)} shows a combination of small (radius $< 1$ $\mu$m) crystalline olivine
(forsterite; labeled `fors') and phyllosilicate (montmorillonite; labeled `phyll') particles.
Panel {\it (b)} shows a combination of small crystalline olivine and large (radius 15 $\mu$m)
glassy olivine particles (labeled `large ol').
The minerals were placed at the same temperature, optimized in the range 500--800 K (as labeled)
and combined with blackbody continua at 1200 and 120 K, as well as the extincted stellar photosphere.
Individual components of the spectral model are shown as dashed lines with labels,
and the model sum is shown as a solid line.
\label{thetamin}}
\end{figure}

First, we consider that the peaks in the feature are due to small grains of specific minerals.
The 9.1 $\mu$m horn is at the approximate location for a phyllosilicate like montmorillonite 
[(Na,Ca)$_{0.33}$(Al,Mg)$_2$(Si$_{4}$O$_{10}$)(OH)$_2$ánH$_2$O] or for amorphous quartz [SiO$_{2}$].
We calculated absorption efficiencies for forsterite, montmorillonite, and quartz using
a continuous distribution of ellipsoids (CDE) and laboratory optical constants 
\citep[from][respectively]{fabian,glotch,henning97}.
 A reasonable fit to the spectrum includes small grains ($<2$ $\mu$m) of
both crystalline forsterite and either montmorillonite or quartz, 
with comparable abundance, together with
continuum emission at temperatures ranging from  700 to 150 K. 
Figure~\ref{thetamin} compares the fit and data.
The silicate feature can
be explained exclusively with crystalline silicates,
though the mineralogy of the larger grains cannot be distinguished.
The crystalline features are emitted predominantly from the hottest part of the disk.
If the luminosity is 40 $L_{\odot}$ (as inferred from the radiative transfer model fit
discussed below), then the crystalline silicates are located about 1.6 AU from the star.
The terrestrial-planet-building materials around the stars IC1396A:$\theta$ and $\beta$
could therefore 
be rather different from most low-mass stars, even in the same globule (cf. Tr37 11-2037
and the other stars in the top of Fig.~\ref{classiispec} that are dominated by amorphous silicates). 

Second, we consider whether grain growth can explain the shape of the silicate feature. 
\citet{honda} showed that if the grains are glassy olivine, the weaker silicate features
can be explained by $\sim 2$ $\mu$m grains and the stronger silicate features can
be explained by $\sim 0.2$ $\mu$m grains. 
To determine whether larger grains can explain the spectrum of IC 1396A:$\theta$, we calculated
absorption efficiencies for a range of grain sizes for glassy olivine \citep{dorschner},
then optimized the disk and size to match the observed spectrum. Figure~\ref{thetamin}(a) shows
the result. As noted previously, the larger grains have a flatter and wider spectra feature,
eventually becoming featureless (and joining the continuum). A peak in the absorption efficiency
of glassy olivine occurs around 8.6 $\mu$m for particles with radii greater than 12 $\mu$m.
As seen in Figure~\ref{thetamin}, the peak is at wavelengths too short to match the observed feature;
however, the general shape is very similar to the observed one, and wider mineralogical search
may yield a better match. 

In summary, the distinctive silicate feature (weak, wide, and `horned') of IC 1396A:$\theta$ and $\beta$ could be due to crystalline mineralogy (with quartz or phyllosilicates
providing the `blue' horn of the feature) or by grain growth (with larger grains providing
the width of the feature and also explaining its weakness).  
For wide samples of T Tauri stars, \citet[][their Fig. 9]{kessler}, 
\citet[][their Fig. 2]{honda}, and \citet[][their Fig. 6 versus their Figs. 3--5]{furlan}
showed that the disks with weaker silicate
features also have `flatter' silicate features, where flatness is defined based on the
ratio of the amplitude of the 11.3 $\mu$m olivine peak relative to the 9.8 $\mu$m amorphous silicate peak. Our results in Figure~\ref{classiispec} also show this trend. 

\citet{furlan} propose an evolutionary sequence
based on the mid-infrared spectral morphology, driven by dust settling and growth.
In this scheme, IC 1396A:$\theta$ and $\beta$ would belong to their `Group D', 
the group with the greatest amount of grain growth.
Other stars with similar spectral shapes in the star cluster include
Tr 37 11-2006 and 24-515 \citep{sicilia07}.
More recently, \citet{kessler07} showed that the `flatness' of the silicate feature is
also correlated with stellar luminosity (i.e. flatter silicate features tend to be
present around more luminous stars). Our data also support this trend. In particular 
IC1396A:$\theta$ and $\beta$ are relatively luminous stars (approximate types A2 and F0) 
with flat features,
and the later-type stars like Tr 37 11-2146 and 21364964+5722270 (K6 and G5) have stronger features.

Another possibility, as discussed above, is that
the mineralogy is different, with increased abundance of phyllosilicates or quartz
around the higher luminosity stars. Phyllosilicates could arise from the
catastrophic disruption of a young asteroid, since aqueously altered silicates are
common compositions of asteroids in particular in the outer main belt 
\citep{rivkin}.
Amorphous SiO$_{2}$ may be produced from catastrophic collisions as well, perhaps
among protoplanets as suggested recently for one star
\citep{inseok}.
It must be acknowledged that low-sensitivity infrared spectroscopy, such as presented
in this paper,
does not uniquely provide 
independent signatures of grain growth or mineralogy, because of the wide range of
possible size distributions and compositions.
High-quality spectroscopy of a wider sample of T Tauri stars will shed
more light on this issue, which is important since the a disk bearing
phyllosilicates versus a disk with moderate amounts of grain growth are such
suggest distinctly different evolutionary state for the disk.

\subsection{Properties of the visible spectra}

\begin{figure}
\epsscale{1}
\plotone{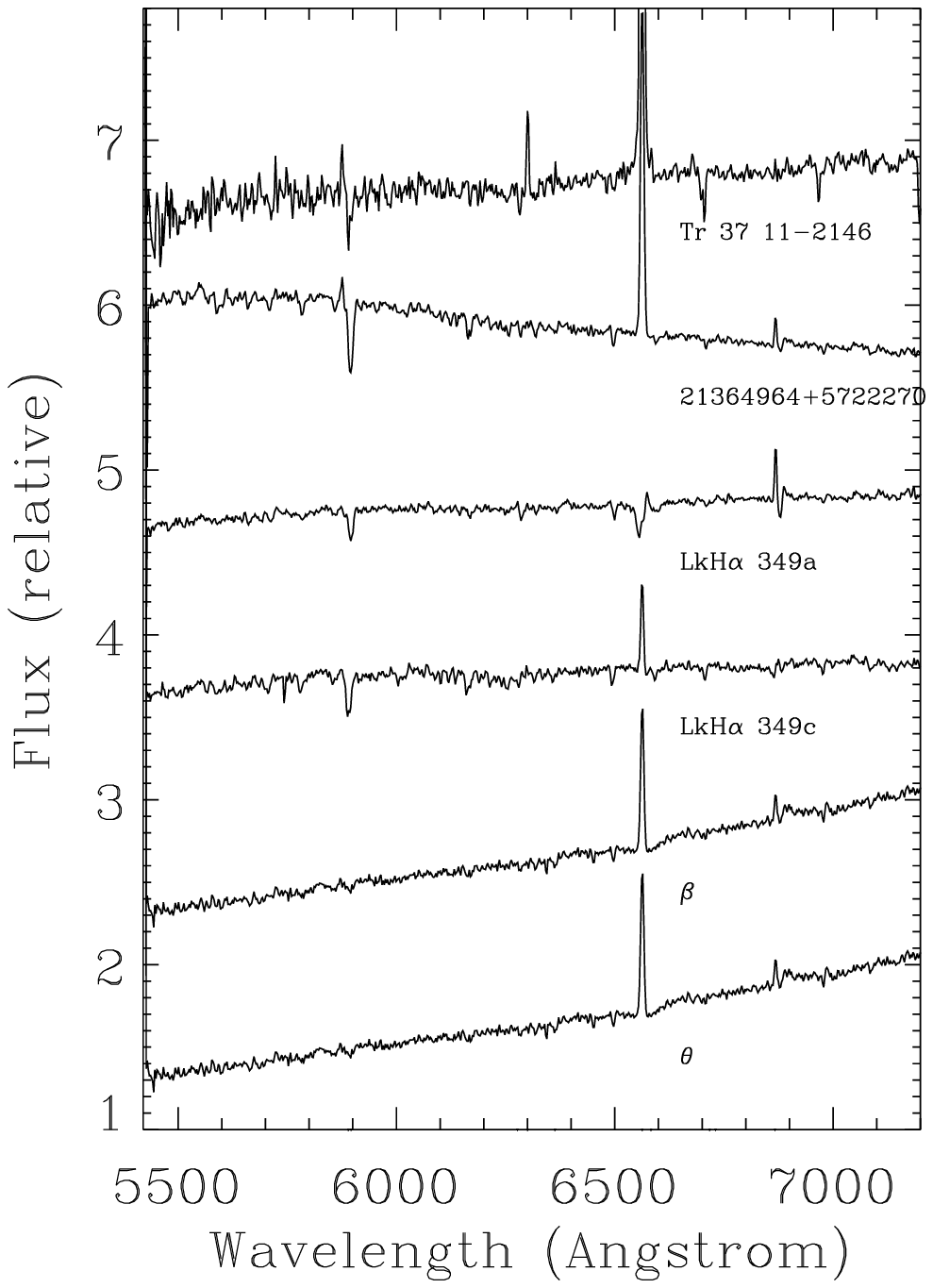}
\epsscale{1}
\figcaption[]{Optical spectra of Class II sources in IC1396A.
Each spectrum was normalized by its median, scaled by 0.8, and shifted by unity from the previous
for clarity. The bright H$\alpha$ lines are evident for all sources, except LkH$\alpha$ 349a, which
has a P Cygni-like line profile as had been seen by Cohen \& Kuhi (1979).
\label{optfig}}
\end{figure}

The optical spectra of the Class II sample show
bright H$\alpha$ emission lines, making them 'classical' T Tauri stars.
Figure ~\ref{optfig} shows some of the spectra.
The stars appear similar to those observed by \citep{sicilia05}, suggesting
they are part of the same population of T Tauri stars widespread throughout
the IC 1396 region, of which IC 1396A is but a small part.
Therefore, we do not elaborate upon their properties here.

\subsection{Association with Whitney protostar models}

The best-matching radiative transfer model from \citet{robitaille}
and \citet{whitney} was identified using the same
procedure as used for the Class I sources (\S\ref{classiwhitney}).
As was found for the Class I sources, the best-matching models
for the Class II sources also had fairly high $\chi^{2}$, so
the models are only to be used as a guide to plausible properties
of the stars. Most of the mid-infrared-selected Class II sources 
are of order solar mass, with moderate-mass and slowly growing disks and
negligible envelope. 
The two known T Tauri stars, LkH$\alpha$ 349a and c are 
fit by 2.2 and 0.7 $M_{\odot}$ stars. The main difference between
them is that LkH$\alpha$ 349c has a massive disk.

One Class II source whose model stands out from the others in terms
of luminosity and mass is IC1396A:$\theta$. 
Its mid-infrared spectrum is very flat, and it could not
be fit by low-mass models as could the others.
The best fit model was a pre-main-sequence A star ($5 M_\odot$).
Such a star has a large predicted photospheric
decrement at 4000 \AA that is not observed in the optical spectrum, and
the predicted V-J=-0.15 whereas the observed V-J=3. 
These could be reconciled if the extinction
$A_{V}>5$ to this star; the best-fitting radiative transfer model has $A_V=8$.
The determination of the stellar mass from the infrared luminosity is of course
very model-dependent; a spectral type in the A--F range are possible.
There are other intermediate mass
stars in Tr 37, including B stars in the IC 1396A region with surrounding nebulosity,
so it is not implausible that IC1396A:$\theta$ is a young, intermediate mass embedded
within a dense clump of material that is not its envelope. These properties plus the
bright H$\alpha$ line make IC1396A:$\theta$ plausibly a Herbig AeBe star \citep{waters}.

The spectral energy distribution of IC 1396A:$\beta$ is also relatively flat in the
mid-infrared, but its 24 $\mu$m flux is not high, which led us to assign
a Class II photometric type. The spectra
presented in this paper confirm that $\beta$ is a T Tauri star and not a
Class I protostar. The best-fitting radiative transfer
model has a large line of sight extinction, $A_V=6.8$, explaining the lack of
bright optical counterpart and infrared-dominated SED. The spectrum of $\beta$ shows
faint PAH features, similar to those seen around other young intermediate-mass
stars \citep{sloan05}, though bright nebular emission in IC 1396A may contaminate
the spectrum at the PAH wavelengths.

The spectral observations of Class II sources show that the mid-infrared-selected 
sample, identified in Paper I, is dominated by highly reddenned, pre-main sequence
stars of a wide range of masses from 0.4--5 $M_\odot$ and
luminosities from 1--40 $L_{\odot}$.

\clearpage

\section{Interaction between the Protostars and Globule}

\subsection{Physical conditions in the globule}

The present configuration of the globule comprises an ionized
layer on the side facing the O star (evident in H$\alpha$ emission), 
a dense bright rim (evident in PAH emission),
and a high-opacity region (evident in extinction and in mm-wave molecular
emission). The dense region comprises a round `head' (centered on LkH$\alpha$ 349)
with a hole in its center, and `trunk' extending to the SW of the head.

Let us denote properties of the ionized layer with subscript $i$.
From radio observations, its emission measure \citep{matthews}
\begin{equation}
EM_i=n_i^2 L_i=2000\,\,{\rm cm}^{-6}\,\,{\rm pc},
\end{equation}
where $L_i\sim 0.07$~pc is the thickness of
the ionized layer; hence we infer 
the density $n_i\sim 160$ cm$^{-3}$
and, assuming temperature $T_i=10^4$ K, pressure
$p_i/k=2\times 10^6$ cm$^{-3}$ K.
The ionized layer is maintained by the O 6.5 star, with
the rate of ionizing photons incident on the surface
of the globule equal to the case B recombination rate
for gas with density $n_i$. 

Let us denote  properties of the globule interior with subscript $g$.
The radius of the head $R_g=0.51$ pc.
From CO and $^{13}$CO measurements, the column density
$N_g\sim 1.5\times 10^{22}$ cm$^{-2}$, and the velocity dispersion
$\delta v_g=2.1$ km~s$^{-1}$ \citep{nakano},
from which we infer the density $n_g\simeq 5000$ cm$^{-3}$ and
pressure $p_g=3\times 10^6$ cm$^{-3}$ K. Thus at present,
the bright rim of the globule is in pressure balance with the
ionizing layer.

However, when the O star first generated its \ion{H}{2} region
and illuminated the globule, the pressure would have been much
higher than that of the ambient gas.
The initial conditions of the medium surrounding the O star are
unknown, but they need not have been exceptional in order to lead
to the present configuration. For example, \citet{gritschm} showed
that an initially randomly turbulent medium evolves into a `pillar'
type morphology, like IC 1396A or the famous pillars of M 16, due
to the pressure and ionization of the O star. The region that became
IC 1396A would have had an initial density higher than average.

After the O star was born ($\sim 4$ Myr ago),
the pressure of the \ion{H}{2} region
drove a shock into the proto-globule. Using the density of the
ionized layer and globule now, and eq. 12-15 of \citet{spitzer},
the shock speed would be $\sim 2$ km~s$^{-1}$. Such a shock
would reach the center of the globule in $3\times 10^5$ yr.
Of course the density of the proto-globule was initially lower, 
and the globule was initially larger. These effects approximately
cancel: the shock velocity $\propto n_0^{-1/2}$, while
for a globule of constant mass the radius $\propto n_0^{-1/3}$, so 
the shock crossing time $\propto n_0^{1/6}$, a very weak dependence
on the initial density. When the shock reached the center of the
globule, a strong over-density developed there. Numerical simulations,
e.g. the landmark ones by \citet{lefloch}, show both the central
concentration and instabilities leading to smaller fragments within
the globule. During the $\sim 10^6$ yr period of compression, 
much of the globule mass becomes ablated in a 
`cometary' phase. IC 1396A is presently in this phase,
with stars of a wide range of mass forming from the over-pressured
globule material.


\subsection{Outflows and Globule Destruction from the Inside}

The central hole in the head of the IC 1396A globule may be only one
of several outflow-related structures in the globule. In the following
subsections we discuss evidence for current outflows in the globule and
for the shaping of the globule by outflows.

\subsubsection{Central cavity and LkH$\alpha$ 349\label{holesec}}

In optical images, a relatively transparent hole, through which background
stars are visible, surrounds the young star centrally located in the
head of the IC 1396A globule. Some wispy nebulosity is present, 
with the appearance of reflection nebulosity.
In molecular line images, the hole appears relatively
vacant of material. \citet{nakano} showed using high-resolution
CO data that the central hole is indeed nearly devoid of molecular
gas. In the {\it Spitzer} images, the hole also appears to be
devoid of material, and furthermore shows a bright rim, with
a particularly well-define `wall' on its W side. The hole size is
$45'' \times 58''$ (EW$\times$NS). The central star, LkH$\alpha$ 349a,
has a spectral type F9e and effective temperature 6165 K
\citep{cohenkuhi,hernandez}. The extinction toward the star is
is nontrivial (despite the nearly empty appearance of the hole).
The extinction is mostly local to IC 1396A, as opposed to
intervening line-of-sight interstellar dust (for which one expects
less than 1 mag extinction at the distance of IC 1396), so
the extinction curve may have high $R_V\sim 5$ typical of
stars behind molecular clouds (as opposed to
the random ISM value of 3.1).
Using the $R_V=5.0$ (3.1) extinction curve, \citet{hernandez} find
LkH$\alpha$ 349a to be a 5.4 (3.0) $M_\odot$ star with luminosity
270 (45) $L_\odot$. The preferred, higher-luminosity solution
is near the birth-line (age $\sim 0.3$ Myr) and consistent
with the other indications of the star's youth such as the H$\alpha$ line,
which has a P Cygni profile indicating a wind with $v_w>300$ km~s$^{-1}$.
The width and shape of the emission and absorption lines suggest
a very high rotation rate $v \sin i \simeq 190$ km~s$^{-1}$, near
the breakup speed; this may be related to its apparently high-speed
wind. We note that LkH$\alpha$ 349a will be a late B- to early A-
type star on the main sequence and is likely a very young Herbig AeBe 
star \citep{herbigrao}, many of which
have similar properties (reflection nebulosity, self-absorbed
H$\alpha$ winds) and are associated with star clusters with age of order 1 Myr
\citep{habart}.

The location of the LkH$\alpha$ 349 inside the hole, and its high luminosity and 
youth, lead one to suspect the star could have cleared the hole.
This can readily be shown to be plausible. If the star produced, for
a time $t_w$, a mechanical luminosity, $L_w$,
leading to a bubble of radius $R_w$, then \citep{castor}
\begin{equation}
R_w = 0.44 \dot{L_w}^{\frac{1}{5}} n_{g4}^{-\frac{1}{5}} t_{w5}^{\frac{3}{5}} 
\,\,{\rm pc},
\end{equation}
where $t_{w5}=t_w/10^5$ yr, $n_{g4}=n_g/10^4$ cm$^{-3}$, and
$L_w$ is in solar luminosities.
If the stellar wind generated the presently-observed hole, 
$R_w=R_h=0.09$ pc, then we can solve for the mechanical luminosity
\begin{equation}
L_w\sim 2 t_{w5}^{-3} \left(\frac{n_g}{5000\,{\rm cm}^{-3}}\right) L_\odot,
\end{equation}
and, for a wind speed $v_w$, the required mass-loss rate 
\begin{equation}
\dot{M}_w=1\times 10^{-7} t_{w5}^3 
\left(\frac{300\,{\rm km~s}^{-1}}{v_w}\right)^2 
\left(\frac{n_g}{5000\,{\rm cm}^{-3}}\right)
M_\odot \,{\rm yr}^{-1}.
\end{equation}
Since $v_w>300$ km~s$^{-1}$,
the mass-loss rate required for LkH$\alpha$ 349a to blow a bubble in
the globule is plausible,
by comparison to other young stars.
Class I sources with luminosity $\sim 10^2$ $L_\odot$ have winds with
$L_w\sim 0.1$--1 $L_\odot$ \citep{spbook}. 
Such a wind would only need to be active
for $\sim 10^5$ yr to provide momentum to sweep the cavity
clear. 
The SED models for LkH$\alpha$ 349a and c indicate ages of 0.4 and 1.2 Myr, 
respectively, allowing time for their wind phase to have passed.
LkH$\alpha$ 349a is rotating exceptionally rapidly
\citep{hessman}, and
has little or no disk, perhaps enabling a relatively spherical and powerful wind.
The star appears close to the stellar birth-line but is
clearly exposed, suggesting that while young, the
star is likely to have completed its main `wind-bearing' Class 0 and I
phases and resides within the bubble that it blew.

The existence of a bubble blown LkH$\alpha$ 349a provides supporting,
circumstantial evidence that it could have been triggered by 
radiative driven implosion. 
If LkH$\alpha$  349a and its bubble had existed {\it before} the formation
of the O 6.5 star, the bubble would not have survived to
this day. The shock front driven by the pressure of the expanding 
\ion{H}{2} region would have reached the center of the globule
long ago, as discussed above. Such a large-scale shock would 
likely have significantly distorted the cavity sent it downwind
of the young LkH$\alpha$ 349a.

\subsubsection{Water maser near IC1396:$\gamma$\label{gammassec}}

Figure~\ref{gamwater} shows water maser emission toward
source IC 1396A:$\gamma$. 
22 GHz H$_{2}$O masers are associated with Class 0 protostars,
and only rarely (4\%) with Class I sources, in a survey of such sources \citep{furuya}. 
The mid-infrared
spectrum of IC1396A:$\gamma$ has the characteristics of a Class I protostar with
a rapidly-forming star accreting material from within a dense envelope.
Is the H$_{2}$O maser from an outflow caught `in transition' from Class 0 to I?
Or is the maser spot from a clump of gas far from the protostar?

As noted by \citet{valdettaro}, the maser spot is significantly offset from
the IC 1396A:$\gamma$ by $5000\pm 200$ AU. The maser peak
coincides with a faint infrared
source in the IRAC images. 
The coordinates of the IRAC source are 
21$^h$36$^{m}$07.8$^{s}$ 57$^{\circ}$26$'$42.8$''$, within $1''$ of
the VLA maser position;
we will refer to this source for convenience as
IC 1396A:$\gamma_{b}$.
The fluxes of $\gamma_{b}$ at 
3.6, 4.5, 5.8, and 8 $\mu$m are 
$<0.2$, $0.6\pm 0.1$, $1.5\pm 0.2$, and $1.2\pm 0.3$ mJy,
respectively; with most of the uncertainty due to blending with the 
much brighter source $\gamma$. 
The corresponding magnitudes are
[3.6]$>$15.3, [4.5]=13.7, [5.8]=12.2, [8.0]=11.8.
No counterpart was evident in the near-infrared images.
By chance, the orientation of one of the mid-infrared (IRS) spectrograph slits covered 
the faint source. The spectrum of $\gamma_{b}$ 
cannot be cleanly separated from that of
$\gamma$, but it is clear that the emission from the source is {\it not} line-dominated.
If the infrared source were a Herbig-Haro type object, where the jet is impacting
a dense clump, then we would expect strong emission lines of H$_2$. 
The IRS short-low spectrum (where the faint source can be separated from $\gamma$)
covers 5.2-14 $\mu$m and would contain the pure rotational H$_{2}$ S(2) through S(7) lines
if they were present. Furthermore, there is no corresponding spot at the
H$_{2}$O maser position in the near-infrared rovibrational H$_{2}$v=1-0 S(1) image.
Therefore, the IRAC source corresponding to the maser is not a shocked 
clump of gas. It is also unlikely to be a background galaxy, based
on the lack of PAH features that would be prominent at 6.2 and 7.7 $\mu$m
modulo some small redshift. The IRS spectrum of the faint source is
dominated by continuum emission, and it looks 
roughly similar to that of the $\gamma$; the difficulty of separating the
sources makes it impossible to study the spectrum in detail from
the available data.

Therefore, it appears that the H$_{2}$O maser is not directly associated with
the luminous infrared source IC 1396A:$\gamma$, nor is it an interaction spot between the outflow and nearby molecular gas. 
The faint mid-infrared source IC 1396A:$\gamma_{b}$ may itself be a protostar powering an outflow responsible for the H$_{2}$O maser. At 24 $\mu$m, 
a very bright source clearly corresponds in position to the
luminous source $\gamma$. There is 
an extension from this source at 24 $\mu$m that corresponds to the faint source seen with IRAC, but it is severely blended:
the offset is $7''$ while the beam size is $5''$. We scaled
 and subtracted the point spread function to
estimate the 24 $\mu$m flux of the fainter source as $\sim 10$ mJy,
corresponding to [24]$\sim$7.1; a very conservative limit is 
[24]$>5.9$. Combining the mid-infrared colors of $\gamma_{b}$ and
comparing to Figure 3 of Paper I (i.e. comparing to the locations
of Taurus and IC 1396A young stellar objects), the H$_{2}$O maser
source $\gamma_{b}$ appears
to be  a low-luminosity Class I protostar. 
This brings
back the original question of why a Class I protostar has an
associated maser. Perhaps this low-luminosity source is a Class I/0
transition object. 

The remarkable result is that
the brightest (only known) H$_{2}$O maser in the globule is
close to the second-most-luminous ($L_{MIR}=1.6 L_\odot$) Class I protostar
in the globule, but it is actually positionally coincident with
a separate, low-luminosity ($0.02 L_\odot$) infrared source.
The separation of 5000 AU is somewhat wider than a binary pair; 
if $\gamma$ has
a mass (star plus envelope) of $\sim 1 M_{\odot}$ then the orbital speed would be 0.5 km~s$^{-1}$,
comparable to the velocity dispersion of the globule.
Class I sources have been found to have a wide distribution of separations, and this
potential pair would be among the widest \citep[cf.][]{connelley}.
There are other low-luminosity infrared sources in this field,
so the total population of protostars may be much larger than the
bright mid-infrared sources identified by eye in Paper I.
Such low-luminosity objects have been found in other nearby
star-forming regions that were surveyed by {\it Spitzer},
including those in dense cores in L1014 \citep{young}, 
L1448 \citep{kauffmann}, and L1521F \citep{bourke};
they have been interpreted as sub-stellar protostars
though the envelopes may contain more than $1 M_{\odot}$.
The prototype of the class, L1014-IRS, was searched for H$_{2}$O masers
on 7 epochs with no detection \citep{shirley}. 
The detection of an H$_{2}$O maser from IC 1396A$\gamma_{b}$
is the first for a low-luminosity object of this class.
Masers luminosities have been found to correlate roughly with the 
bolometric luminosity of their driving protostars \citep{furuya,shirley},
with $L_{H_{2}O}=3\times 10^{-9}L_{\odot}$ and over an
order of magnitude of scatter. One could simply use this scaling to 
estimate or constrain the luminosity of the source, despite the large scatter
and wide range of variability (a factor of 180). For IC 1396A:$\gamma_{b}$,
we would obtain $L_{bol}\sim 0.1$--$1 L_{\odot}$ depending which epochs
are averaged together. Such a luminosity is not inconsistent with the
observed mid-infrared luminosity, though it does suggest that the much
of the luminosity emerges in the far-infrared. 

\subsubsection{Molecular Outflow from IC 1396A:$\gamma$}

\begin{figure}
\epsscale{1}
\plotone{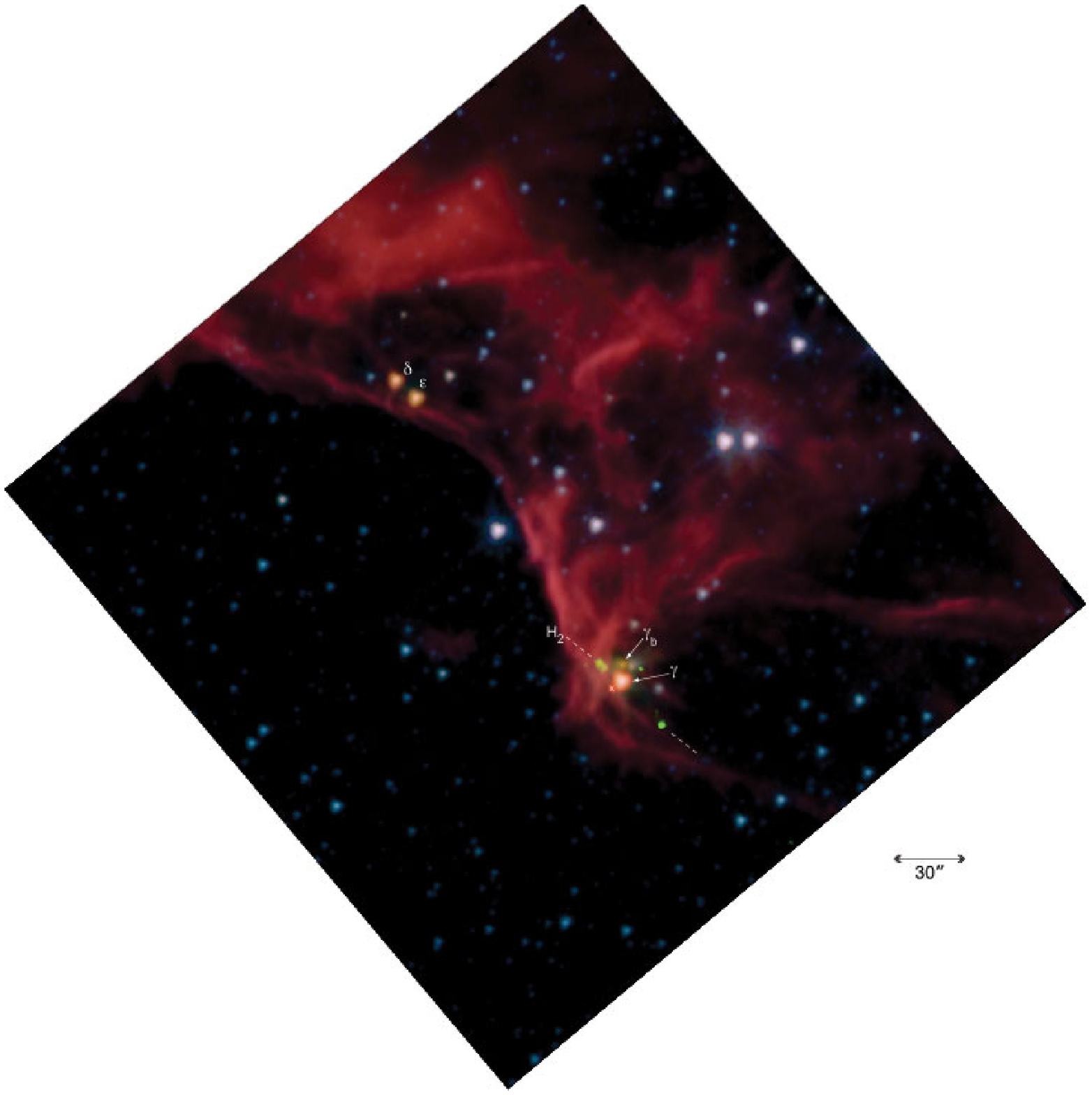}
\epsscale{1}
\figcaption{{\it Spitzer}/IRAC color image of the region around
the protostar IC 1396A:$\gamma$. The 3.6 $\mu$m image is coded as blue,
4.5 $\mu$m as green, 5.8 $\mu$m as orange, and 8 $\mu$m as red.
The Palomar/WIRC H$_{2}$ 2.12 $\mu$m image is coded as light green.
Protostars IC 1396A:$\gamma$, $\delta$ and $\epsilon$ are labeled.
Source $\gamma_{b}$ is coincident with the 22 GHz H$_{2}$O maser. 
A small `$\times$' indicates an IRAC image artifact exactly 4.88$''$ 
(4 IRAC pixels) SE of $\gamma$. Dashed lines indicate the two shocked H$_{2}$
blobs on either side of source $\gamma$. 
The scale bar is $30''$=0.11 pc=0.36 l.y.=22,000 AU long.
\label{gammazoom}}
\end{figure}

The Palomar 2.12 $\mu$m H$_{2}$1--0 S(1) continuum-subtracted image shows two knots of
emission near IC 1396A:$\gamma$. Figure~\ref{gammazoom} shows that these knots are located
diametrically opposite each other along a line that passes through $\gamma$, a 
configuration strongly suggestive of a jet, powered by $\gamma$, that is striking molecular
knots forming dense Herbig-Haro objects \citep{hhobject}.
The short length of the outflow (in particular compared to pc-scale outflows)
is likely due to its propagation into dense walls of
the globule.
That the outflow is seen in H$_{2}$ but not other tracers is also likely due to
the dense material the jet interacts with.

\def\extra{
SMT observations...
fitting to gamma CO(3-2) line
 Line      Area               Position           Width              Intensity
  1   26.329     (  0.270)   -8.179 (  0.011)    2.348 (  0.026)   10.534    
  2   30.999     (  0.489)   -8.842 (  0.059)    7.953 (  0.088)   3.6619    
alpha
 Line      Area               Position           Width              Intensity
  1   29.772     (  0.155)   -8.078 (  0.005)    2.025 (  0.012)   13.809    

12M-HCO+
gamma
 Line      Area               Position           Width              Intensity
  1   3.7255     (  0.198)   -8.174 (  0.063)    2.358 (  0.139)   1.4843    
}

\begin{figure}
\plotone{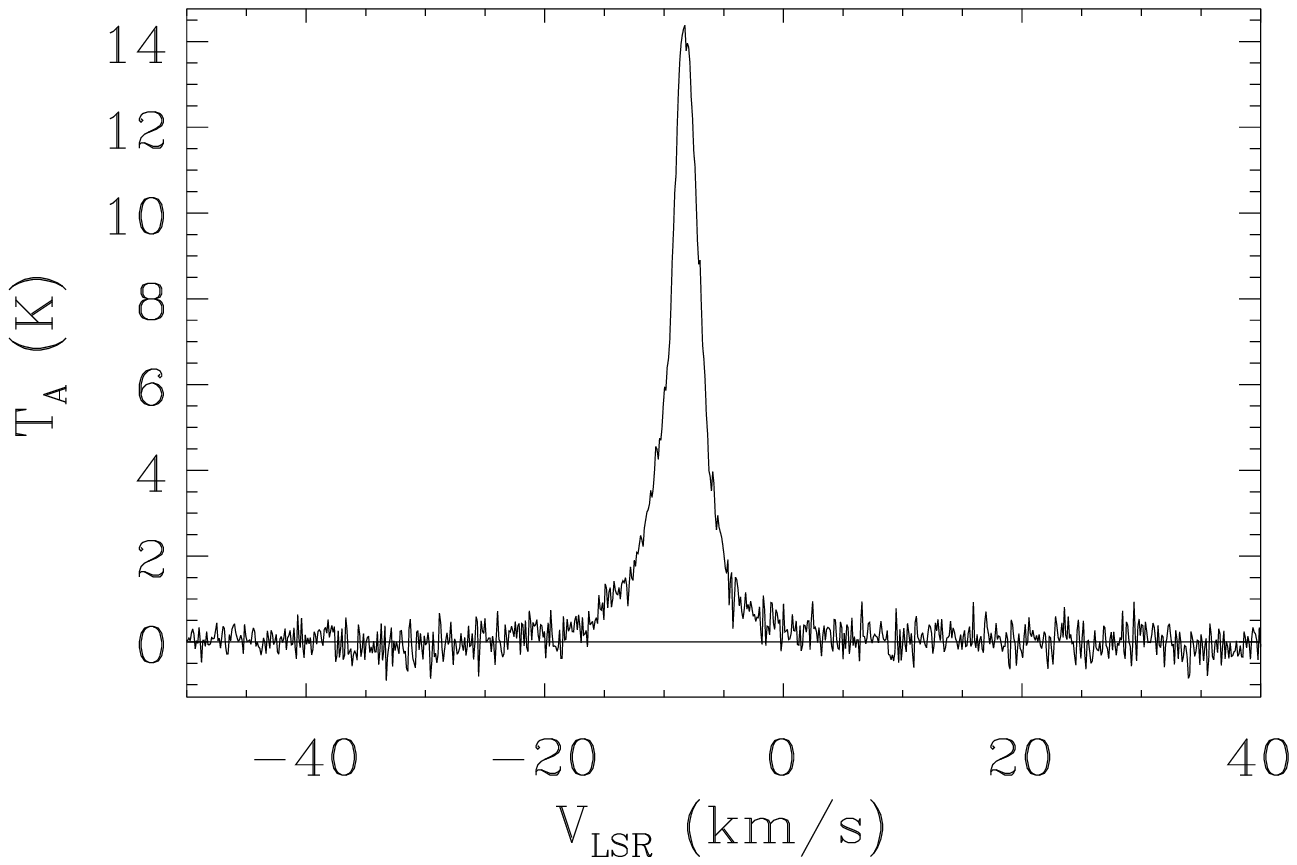}
\figcaption{CO(3--2) spectrum of the protostar IC 1396A:$\gamma$. 
\label{gammaco32}}
\end{figure}

Further evidence for a molecular outflow is found from observations this 
source using the Arizona Radio Astronomy Observatory Submillimeter
Telescope in the CO(3--2) line ($23''$ beam, including both $\gamma$ and $\gamma_{b}$.
Figure~\ref{gammaco32} shows
a prominent blue wing; a two-component fit of 
narrow line typical of the globule plus a second Gaussian with a full width at half maximum
of 7.9 km~s$^{-1}$ and peak antenna temperature 3.7 K. 

\clearpage

The presence of outflows from young stellar objects in IC 1396 is not unique
to the globule studied in this paper. \citet{ic1396n} found HH objects and a 
0.6-pc scale outflow in IC 1396N. The find that HH 777 emerges from the dense
globule into the \ion{H}{2} region. In the following section we discuss the
potential influence of outflows on IC 1396A.

\subsection{Compartments within the globule due to outflows}

The internal structure of the globule around IC 1396:$\gamma$ (Figure~\ref{gammazoom})
is complicated, but it can be described
as follows. There is a bright `wall' located directly $\sim 10''$ upstream 
(i.e. toward the central O star of the \ion{H}{2} region) of
$\gamma$, with a curved shape that appears to merge with two east-west filaments
that pass $\sim 30''$ north and south of $\gamma$; the filaments
continue downstream of $\gamma$ for $\sim 180'$ before merging with
other structures in the globule. 

Using the equations above, assuming the upstream wall is a boundary
between the wind and the turbulent pressure of the globule (or the
ionized layer surrounding the globule, which has comparable pressure),
the required mechanical luminosity of the wind
$\sim 0.02 L_\odot$ and the mass-loss rate $\sim 10^{-7} M_\odot$~yr$^{-1}$
(assuming an outflow speed $\sim 30$ km~s$^{-1}$ typical of such outflows)
if the wind has been active for $10^5$ yr. Such a luminosity is plausible,
as $\gamma$ is a class I/0 source with luminosity $\sim 30 L_\odot$, for
which outflow mechanical luminosities $\sim 0.1 L_\odot$ are typical
\citep{spbook}. The shape of the cavity around $\gamma$, if it is wind-blown,
suggests a significantly lower density on its downstream side. If we naively
use the wind bubble radius equation for the upstream and downstream 
directions separately, then with a size some 18 times longer in the downstream
direction, the density would be lower by a factor of $10^6$, meaning the
region downstream from $\gamma$ was already essentially empty, with 
$n<1$ cm$^{-3}$, compared to the rest of the globule. Thus it seems unlikely
that the structure around $\gamma$ is an asymmetric, wind-blown bubble.
Rather, its appearance and physical properties suggest a wind blowing into 
a high-pressure front of dense globule material upstream, and expanding
relatively freely downstream into a region of much lower density. 

Could this shape be 
a wind bubble blown by $\gamma$ and other nearby protostars? 
We can use the same type of estimate
as in the previous section to estimate the wind properties. The globule
is very optically dark near $\gamma$ indicating a high column density.
The H$_2$O maser line is narrow, but we cannot use this as an estimate of
the wind speed because masers are highly localized spots within the outflow.
Instead, a lower limit to the molecular outflow speed is obtained from
the range of velocities observed for the H$_2$O maser, which spanned 5 km~s$^{-1}$
over the observed epochs (Fig.~\ref{gamwater}). 
The actual outflow velocity is likely to be much higher
depending on projection effects (as the masers are only individual spots
in the outflow and can only under-represent the true range of velocities)
and the bias that the H$_2$O emission traces only dense, warm molecular
gas and not the likely ionized and much faster material closer to the star.

This empirical reasoning can be further extended to the rest of the globule. 
There are 5 prominent `voids' within the globule and its main SE extension.
All of the bright Class I sources for which we obtained IRS spectra are
located within voids. The one containing $\gamma$ was just discussed in
the previous paragraph. The one containing $\delta$ and $\epsilon$ is
similar in many ways to the one containing $\gamma$, in size, shape,
and orientation. Since $\delta$ and $\epsilon$ are so close, it is possible
that their winds combine to form this void. Three large voids are located within 
the globule head. These include the one containing Lk$\alpha$ that was
discussed above, as well as one that contains protostar $\alpha$. The 
third  contains sources $\mu$ and $\zeta$, with its E edge overlapping the globule
edge; however, its NE boundary is clearly not the globule edge and has a
curvature consistent with a roughly parabolic surface with $\mu$ at its
focus and $\zeta$ near its edge. 

Since the major voids all contain either mid-infrared-bright, Class I 
protostars, or the Herbig AeBe star in the globule hole, we suggest that
the globule is being reshaped from the inside, with each protostar residing
in a `compartment' that it has blown via its outflow. Turning the argument
around, we should see what influence each of the photometrically-defined
Class I sources may have had on its environment. For this we consider
the sources in the Class I region of the color-color diagram from 
\citet{reachIC}. In order of 8 $\mu$m flux, these are $\alpha$,
$\gamma$, $\delta$, $\epsilon$, $\eta$, $\zeta$, $\mu$. 
All except $\eta$ are accounted for in voids. Since $\eta$ is located in a
relatively busy portion of the globule, it would be difficult to separate
any potential cavity from other overlapping structures. 
There are other, poorly-defined or smaller voids within the globule.
One such void in the SW `trunk' extension is located just W of $\delta$+$\epsilon$ 
and just N of $\gamma$; this void is not as well defined and contains no 
mid-infrared source. 
Thus it appears
that all (but one) of the Class I protostars are located within voids,
and all (but one) of the voids contain Class I protostars or the
Herbig AeBe star. 

The evidence suggests that protostellar outflows are shaping the globule from
the inside, while the pressure from the \ion{H}{2} region maintains the
bright rim in pressure equilibrium. The outflows may have a double
significance. Not only are they destroying the globule's
structural integrity, leading to its further fragmentation and dissipation,
but they may also drive and maintain turbulence, which would otherwise dissipate 
quickly in the dense gas, within the globule \citep{linak06}. This 
induced turbulence may be what prevents the entire globule from collapsing 
into stars, limiting the star formation efficiency.

\subsection{Proplyd-like objects near IC 1396A}

In addition to the protostars and wind signatures, the mid-infrared images
contain features indicative of dense, starless cores.
The term `proplyd' was coined by \citet{odell} to refer to protoplanetary
disks made observable as silhouettes in HST images of 
the Orion \ion{H}{2} region:
a generic description would include a central star, an elliptical, dark
patch, and a bright rim. 
Without a central star, we have a proplyd-like 
object that could be starless or harbor a protostar. 
There are numerous
dark globules within IC 1396, ranging from the relatively large
and bright ones like IC 1396A or IC 1396N \citep{ic1396n}
to smaller fragments dispersed throughout
the \ion{H}{2} region. We mention here some of the smallest ones, 
because the large, bright-rimmed globules contain multiple embedded 
infrared sources and do not meet the spirit of the definition as
being produced by the circumestellar material of a single young star.

In the $10^\prime$ region around IC 1396A, there are several small clouds that have
distinct {\it dark cores} surrounded by {\it bright rims} and often with
cometary tails. These look similar to the proplyd-like objects 
in other \ion{H}{2} regions. \citet{demarco} studied those in 8
regions, finding that only one objects in M~17 contains a central
star and may be a real proplyd (joining those originally found in 
Orion). \citet{smith} explains the paucity of such objects outside
Orion as being due to the short lifetime of the proplyd phenomenon.
(Only Orion contains numerous stars $< 0.5$ Myr very close to the
O stars.) 
Are the proplyd-like objects in IC 1396 and other \ion{H}{2}
regions small, individual-star-forming globules, or are they
transient molecular cloud fragments en route to photoevaporation?

\begin{figure}
\epsscale{.23}
\plotone{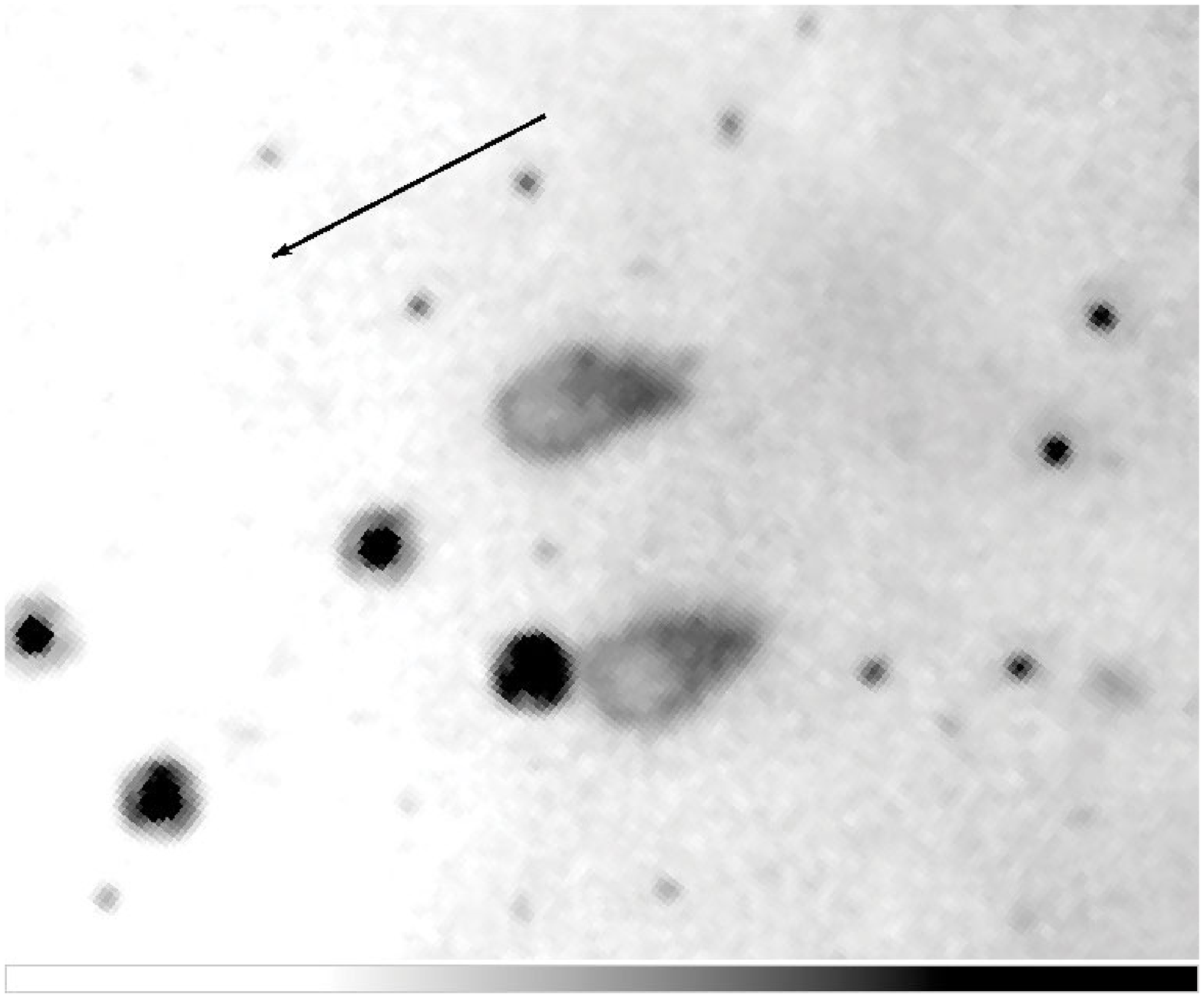}
\plotone{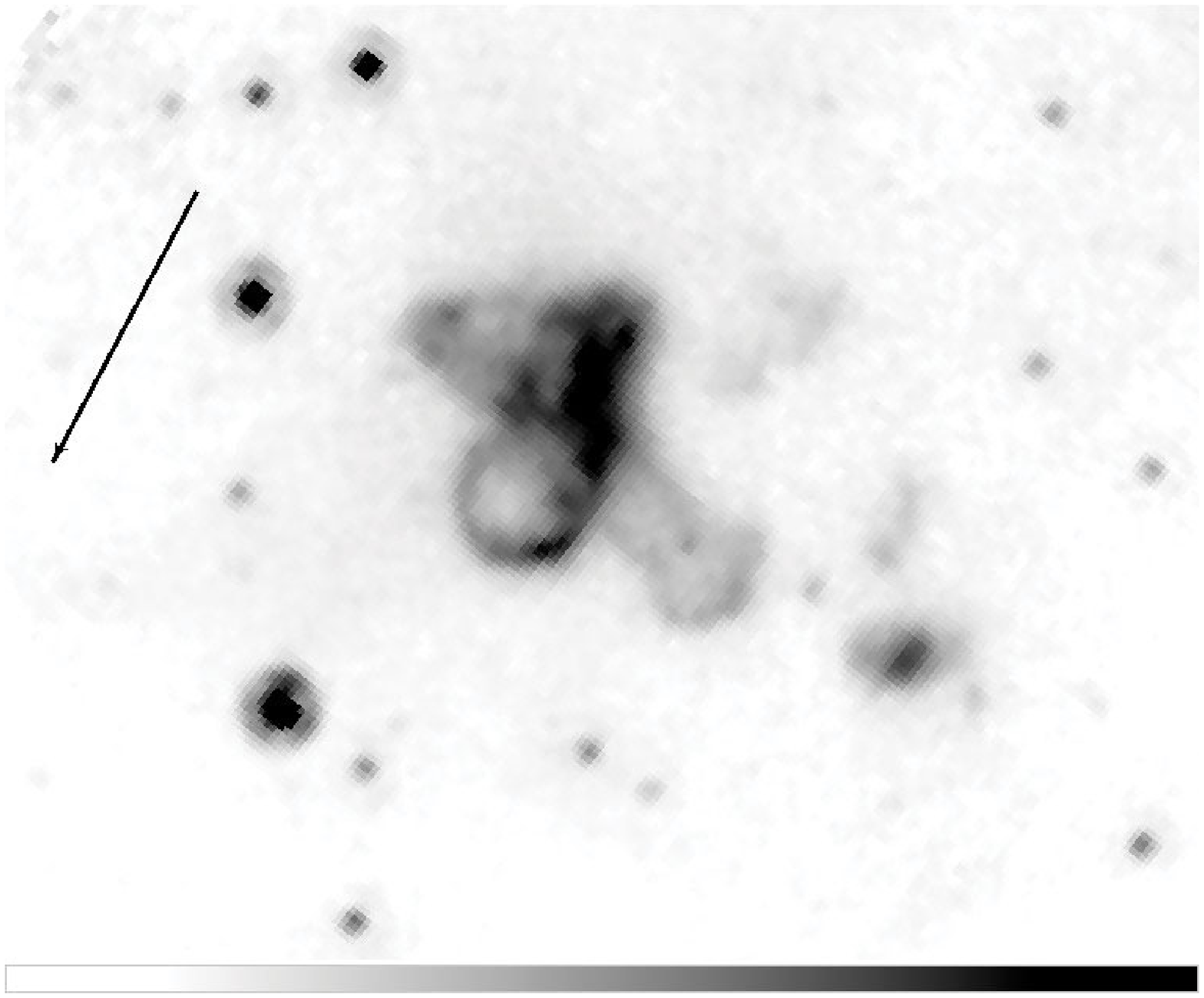}
\plotone{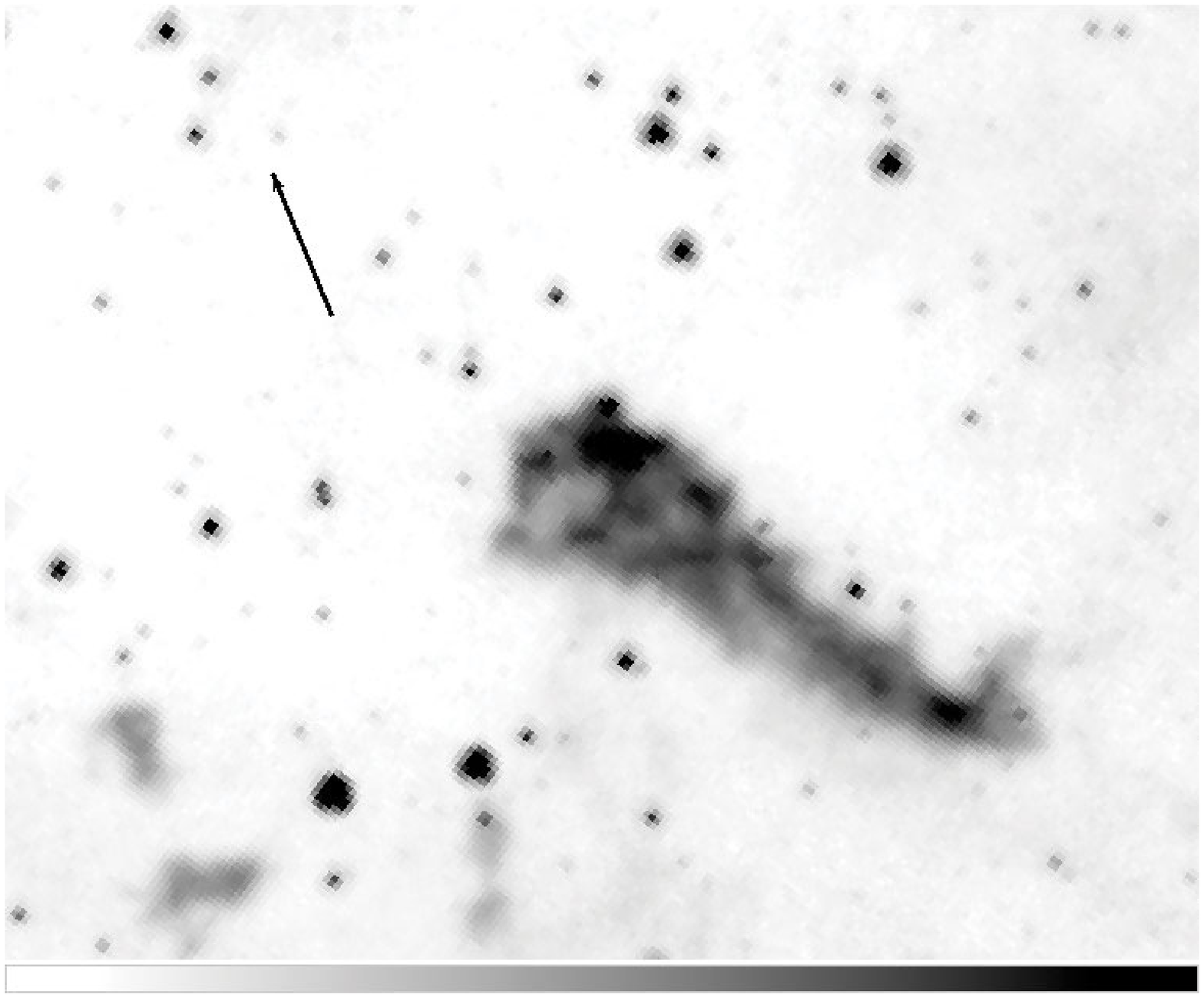}
\plotone{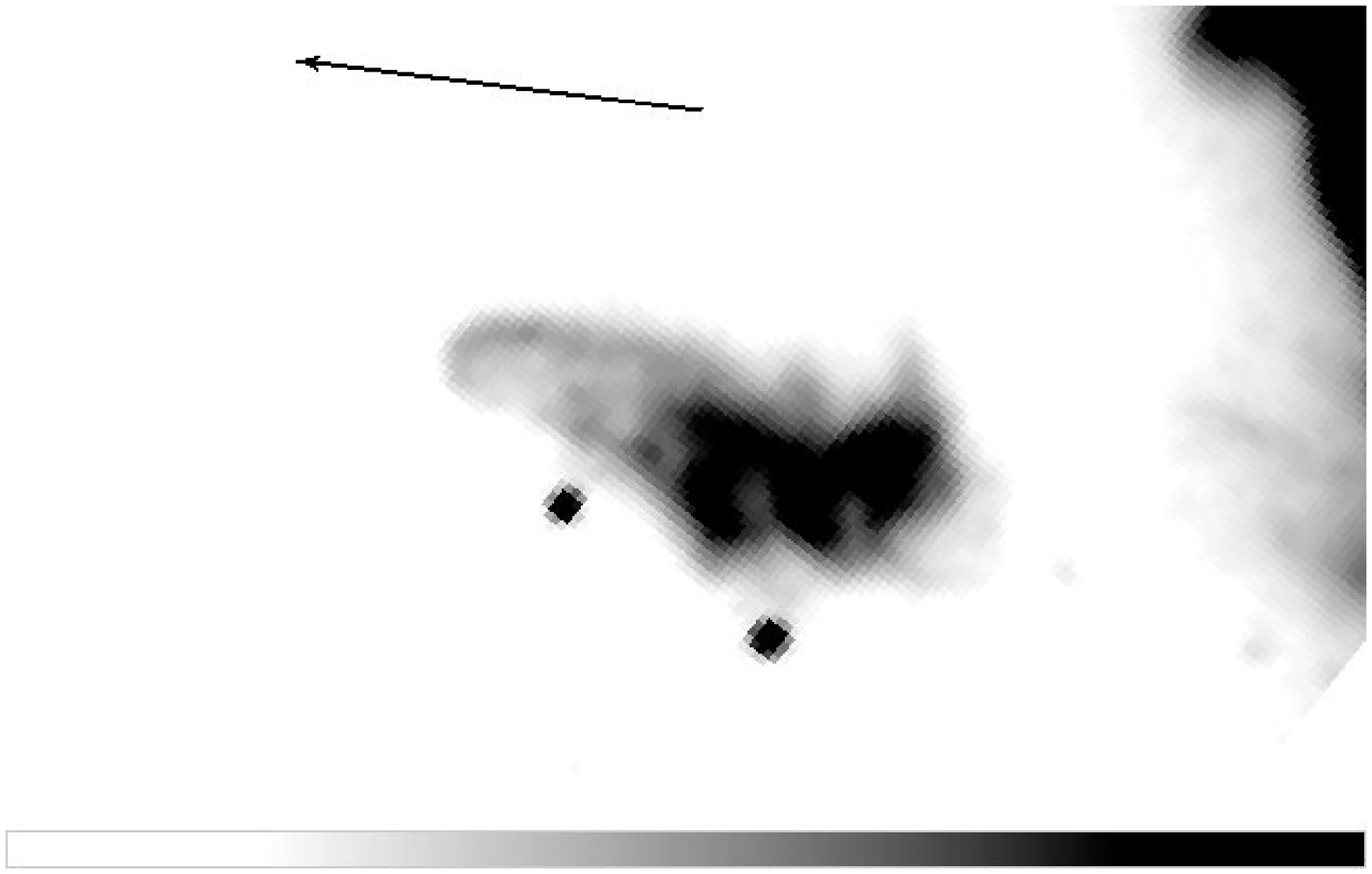}
\epsscale{1}
\figcaption{{\it Spitzer} 8 $\mu$m images of proplyd-like objects in
IC 1396. Coordinates are give in Table~\ref{proplydtab}; panel
{\it (a)} contains objects A and B; 
panel {\it (b)} contains object C; 
panel {\it (c)} contains object D;
and panel {\it (d)} contains object E.
An arrow in each panel shows a $30''$ long vector pointed toward the
central O star HD 206267. All images are oriented with N up and E to the left.
\label{proplydfig}}
\end{figure}

Figure~\ref{proplydfig} shows images of 5 proplyd-like objects in
IC 1396, and Table~\ref{proplydtab} tabulates their properties.
For objects A, B, C, and E, the position angle of the vector from
the object to the central O star (O* PA) is nearly equal to
the position angle of the vector from the tail to the head
(Tail PA), indicating the tails are material ablated from the
head, as is the case on larger scale for the IC 1396A globule.
The central holes are marginally resolved with IRAC at 8 $\mu$m, 
having diameters 5.0 to 15$''$ compared to the PSF diameter of 2.4$''$.
For each object there is a distinct, dark `head' that could be
due either to a lack of material (i.e. an empty hole) or 
extinction. If we attribute the `heads' as being dark due
to extinction, then we
require an optical depth at 8 $\mu$m $\tau_8>1$, which
corresponds to $A_V>10$ and $N_H\sim 1.8 \times 10^{22}$ cm$^{-2}$.
The heads are not resolved in the MIPS images; the 24 $\mu$m image
is consistent in morphology with a suitably smoothed and resampled
8 $\mu$m image (so the 24 $\mu$m morphology is consistent
with either extinction or a central hole).
The central volume density required in the heads are of order 
$10^6$ cm$^{-3}$
and masses of order 0.1 $M_\odot$. The free-fall time is short
($10^5$ yr) in such high-density gas, but the cores can be supported
if they have turbulent velocities around 0.2 km~s$^{-2}$ or
a significant magnetic field. The properties of the heads are
typical of small dense cores in nearby star-forming regions \citep{myersref}.
The analysis just presented is similar in nature to that of
\citet{demarco}, who set the optical depth to H$\alpha$ emission
to of order unity and concluded their proplyd-like objects are
low-density and transient. In contrast, we find significant optical
depth at 8 $\mu$m, requiring a significantly higher density. The
near balance between likely turbulent and magnetic pressure versus
gravitational force suggests the cores {\it may} indeed be forming
stars. Their dynamical configuration, with a central density enhancement,
bright rim, and ablated tail, suggest that the central core density
could be higher than indicated at the angular resolution presented here,
in which case that portion of the dense core will almost certainly
collapse into a small star.

An alternative explanation for the proplyd-like objects is a standoff
between a wind and the ionization front. In this case the deficit in
brightness at the `head' is due at least partially to a lack of
material, instead of extinction. Using the same relations as 
in \S\ref{holesec}, the required wind mechanical luminosity 
$L_w=0.002$--0.09 $t_{w5}^{-3}$ $L_\odot$. An outflow power of this order
is typical for protostellar sources with luminosity or order 1 $L_\odot$.
If the luminosity of such a source were emitted predominantly at
24 (70) $\mu$m, its flux would be of $\sim 0.5$ (1) Jy, which is clearly
ruled out with the {\it Spitzer} data. The observed properties of the
proplyd-like objects is therefore only consistent with being wind
cavities if the embedded sources are significantly under-luminous
(in photons) or high-powered (in wind). With the available information,
we suggest the hypothesis previously mentioned, of an evaporating dense
globule, is more likely.

While we described the objects as `proplyd-like' based on analogy between
their morphology and those of the silhouette protoplanetary disks in the
HST image of Orion,
another analogy is the evaporating gaseous globules (`EGGs') that are
seen in many star-forming regions \citep{hesteregg}. The primary physical
difference between EGG and proplyd
is the lack of a bright central star, although
even some EGGs are found to have low-luminosity stars or even outflows
when observed at very high resolution. A point of distinction between
these EGG-like objects is that they are not connected to a larger `pillar'
or globule that is being eroded by the ionization front. Instead, the objects
we present here are isolated. 
They were most likely small regions of 
relatively high density that are currently being illuminated by the
star, just like miniature versions of the IC 1396A globule. 
The proplyd-like objects discussed here are similar to the 
Thackeray's globules in IC 2944, which appear to be starless \citep{thackeray}.
That these
small globules lack indications of star formation in the {\it Spitzer}
images shows that the mechanism of triggered star formation may not
operate on the smallest globules, which may be evaporating so quickly
(perhaps due to their larger surface area to mass ratio) that stars to
not have time to form in their cores.

\clearpage

\clearpage

\section{Star formation in IC 1396A}

\subsection{Pre-Main Sequence ages for infrared-bright stars near IC 1396A}

\begin{figure}
\epsscale{1}
\plotone{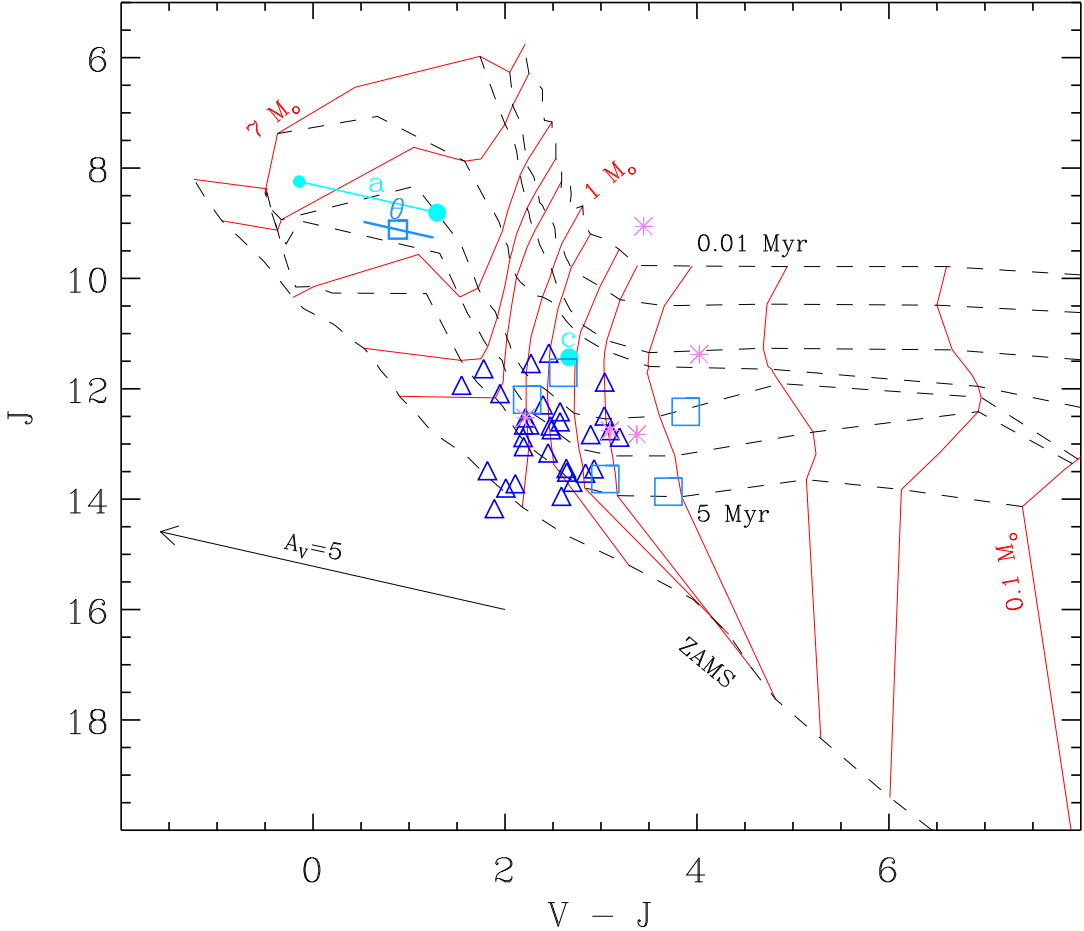}
\epsscale{1}
\figcaption{
Hertzprung-Russell diagram for IC 1396A, using theoretical 
models from \citep{siess} with $Z=0.04$.
Solid red lines are pre-main sequence Hayashi
tracks for stars with masses (right to left)
0.1, 0.2, 0.3, 0.4, 0.5, 0.6, 0.8, 1, 1.3, 1.5,
1.9, 2.7, 5, and 7 $M_\odot$.
The first 7 dashed black lines 
show isochrones for stars with ages 
(top to bottom)  
0.01, 0.03, 0.1, 0.3, 1, 2, and 5 Myr;
the lowest dashed line shows the zero-age main sequence (ZAMS). 
Two filled cyan circles linked by a cyan line labeled `a'
show the location of LkH$\alpha$ 349a for extinction curves with
$R_V=3.1$ and 5; the larger circle is the preferred, $R_V=5$ result.
A separate cyan filled circle labeled `c' marks the location of LkH$\alpha$ 349c, on the 0.5 Myr isochrone and 0.6--0.8 $M_{\odot}$
Hayashi track.
Blue triangles
mark the locations of low-mass pre-main sequence stars within 
the IC 1396A field \citep{sicilia05}.
DodgerBlue-colored squares show the locations of the
mid-infrared-selected (and optically detected) stars
from Table~\ref{opttab};
IC 1396A:$\theta$ is labeled and includes a line for a 1 mag
range of extinctions.
Violet asterisks mark the locations of stars just upstream of the bright rim 
of IC 1396A from Table~\ref{rimfront}.
\label{siessplot}}
\end{figure}

To determine the ages of the pre-main sequence stars near IC 1396A,
we used the models from \citet{siess} for a relatively high 
metallicity $Z=0.04$, based on the assumption that the 
young Cep OB2 region should be more enriched 
than the cloud from which the Sun formed some 5 Gyr ago.
Figure~\ref{siessplot} shows the location on a color-magnitude
diagram of the low-mass pre-main sequence stars within a 
$9'$ radius of IC 1396A from 
\citet{sicilia05} and from this paper.
The photometry and extinction correction are not accurately known, 
so the large spread in H-R diagram and the precise locations of
individual stars should not be over-interpreted.
Taken at face value, LkH$\alpha$c is very young ($\sim 0.5$ Myr)
and possibly coeval with the more-massive LkH$\alpha$ 349a 
(discussed in \S\ref{holesec}).
The location of the source IC 1396A:$\theta$ can be only roughly 
estimated since it is not detected at 5500\AA; for illustration
we placed it at the location appropriate for V=R+.5=17.5.
The mid-infrared sources tend to be younger than the optically-selected stars.
Most of the stars are in the 1--5 Myr range,
including the infrared-selected Class II sources for
which Palomar and {\it Spitzer} spectra were obtained.


\subsection{Generations of Star Formation in IC 1396A}

Apart from LkH$\alpha$ 349, which formed near the center of the globule,
the triggered star formation in smaller condensations may be expected to follow a 
sequence. For example, \citet{sugitani95} found an age gradient such that
T Tauri stars lie closer to the triggering O star than {\it IRAS} sources,
which tend to lie behind the bright rim. \citet{matsuyanagi} showed this
in detail for an individual bright rim.
Does a similar sequence obtain among
the young, mid-infrared sources in IC 1396A? 
There is a handful of Class I sources within IC 1396A---more than the typically 1 {\it IRAS} source
for bright rims where the sequential trend was first noted
\citep{sugitani95}.
The youngest protostars in IC 1396A are likely
those with the most `extreme' infrared colors; these
are $\epsilon$, $\delta$, $\gamma$, $\alpha$, in order of increasing [8]-[24]
(and [3.6]-[5.8]) color. 
The sources $\epsilon$, $\delta$, and $\gamma$ are located just behind (0.02 pc) bright rims.
On a larger scale, all of these protostars are located downstream of the center of the
globule, and all but $\alpha$ are in the `trunk' portion, known as SFO 36, 
which may have been
swept back from the `head' of the globule. 
The luminous Class I sources
are further behind the bright rim than LkH$\alpha$ 349 and
the hole it has carved from the center of the head of the globule. 
Thus it appears that the centroid 
location of the youngest protostars
is significantly behind the main concentration of globule mass.

Inspired by this observation, we searched for Class II young stellar objects just
upstream from the bright rim. Indeed, there is such a set of infrared 
sources. We measured the mid-infrared fluxes of several of the 
24 $\mu$m sources, and
found the corresponding entries in 2MASS ($J$, $H$, $K$) and where possible in
the tables by \citep{sicilia05} (optical magnitudes, spectral type)
or the USNO-A2 catalog.\footnote{From the USNOFS Image and Catalog Archive
   operated by the United States Naval Observatory, Flagstaff Station
   (http://www.nofs.navy.mil/data/fchpix/).}
Table~\ref{rimfront} lists the  4 sources with [8]<11.
All of these sources are visibly exposed, despite being mid-infrared 
selected. For the 3 stars with spectra from \citet{sicilia05}, 
emission lines make them classical T Tauri stars, with ages $>10^6$ yr.
A detailed study of such sources these sources is not possible without
spectroscopy, because of variable extinction.
One of the sources has essentially stellar color in the mid-infrared,
but the optical counterpart is very faint (though it does have
measured proper motion in the USNO catalog, indicating it is
clearly not a background galaxy).

\begin{figure}
\plotone{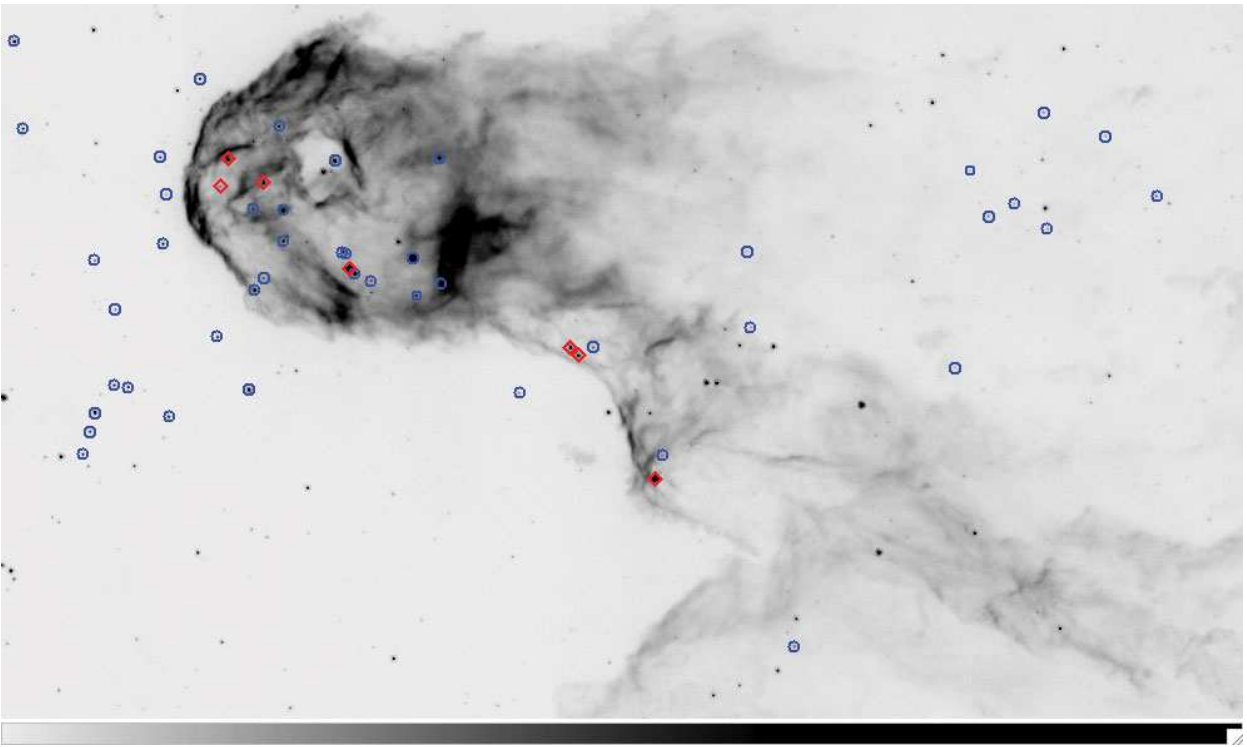}
\figcaption{{\it Spitzer} 8 $\mu$m image of IC 1396A with sources
photometrically identified as Class I enclosed in red diamonds, and
those identified as Class II enclosed in blue circles.
\label{class12}}
\end{figure}

Figure~\ref{class12} shows the distribution of sources photometrically 
identified as Class I and Class II using the {\it Spitzer}/IRAC and
MIPS data from Paper I.
The Class II sources are distributed throughout the region, while the
Class I sources are clearly located within the globule. There may be a
higher density of Class II sources toward the globule head and just upstream,
though the number of sources is too small to determine if the over-density is
significant. 

Note that we do not see the same trend of locations of 
{\it near}-infrared sources relative to mid-infrared that \citet{sugitani95}
found for some other bright rims. Whereas they found the sources
with $J-K>1.2$ are located preferentially upstream of the bright rim,
we find instead that sources with $J-K>1.2$, and also even redder stars
with $J-K>1.8$, tend to be located 
preferentially toward the globule. However, we do not believe this is a
true association of large number of Class II YSOs with the densest part
of the globule. 
Instead the distribution appears more consistent with a pattern of {\it extinction}
of background sources, plus the Class I sources that are associated with the globule.
To show this, we mapped the stars with
more normal colors ($J-K<1.2$), finding a fairly uniform distribution over
nearly the entire region except toward the globule, which appears
as a dramatic deficit of blue sources. In contrast, the distribution
of red sources traces the globule accurately, and the number of red
sources is exactly what is needed to `fill in'
the missing sources toward the globule. In \S\ref{extinctsec} we described
the extinction map generated from near-infrared source colors; 
this extinction map is very closely related to the 
distribution of CO and diffuse infrared emission. Similar
extinction maps were obtained for IC 1396 globules by \citet{froebrich}.
Therefore, it would be challenging to use a simple near infrared 
color criterion to select near-infrared sources 
as T Tauri candidates for IC 1396A.

With the available data, we can
summarize that there are some mid-infrared-bright sources just upstream from
the IC1396A bright rim, and most of these sources may be classical T Tauri
stars. A small minority of the upstream stars may be  highly 
embedded sources, but they are {\it much fainter} and less numerous
than the Class I sources in the globule. 
There is a much wider distribution of Class II sources throughout the entire,
large IC 1396 region. 
Thus we see some evidence for
an age gradient versus density: the protostars are associated with the
dense gas and the Class II objects are dispersed.
Two possible mechanisms have been suggested for age gradients of this sort, 
and both of these mechanisms can work in concert.
Stars that form earliest are gradually exposed as the
pillar recedes due to progress of the ionization effect and `rocket'
acceleration in response to the ionized evaporative flow toward the 
exciting O star. While this scenario does appear to match the
configuration for IC 1396A, in this scenario the young stars are expected to
appear near the `tips' of pillars \citep{hester05}, whereas we observe
the presently-forming stars further downstream. 
The other scenario is small-scale sequential
star formation \citep{lefloch,sugitani95}, where the earliest stars to form in
a globule actually contribute to the collapse and triggering of
the next generation. 

\section{Conclusions}

We studied a mid-infrared-selected sample of young stellar objects from {\it Spitzer} 
imaging (Paper I) using spectroscopic observations from radio
to optical wavelengths. The photometric classifications in Paper I are confirmed by
the spectroscopy. The Class I sources have deep silicate absorption features and
the vast majority of their luminosity arises in the infrared. None of them show
H$_{2}$O masers (characteristic of Class 0) or optical (Class II) counterparts. 
The only H$_{2}$O maser associated with our original sample turns out to be a
separate object, IC 1396A:$\gamma$b, with much lower luminosity $\sim 0.02 L_{\odot}$
than the protostar we identified originally, IC 1396A$:\gamma$.
The mid-infrared-selected Class I sources can be fitted with 0.1--2 $M_{\odot}$ stars
with ages $<2\times 10^{5}$ yr. 

Compared to stars forming in the more quiescent Taurus molecular cloud,
the physical conditions of the
globule from which the stars are forming include a higher radiation field
due to the O6 star exciting the \ion{H}{2} region and nearby young B stars.
There is also a higher pressure due to compression by the ionization front from 
the O6 star and winds from the many young and still-forming stars. The wind
from the $\sim 5 M_{\odot}$ star LkH$\alpha$ 349a has cleared a hole in the center
of the dense globule, and outflows from the Spitzer-discovered Class I objects
appear to have swept material into `compartments' whose walls are over-pressured.
The spectroscopy results show that the Class I objects are brighter in the 
far-infrared than the theoretical models
predict, suggesting they have a warmer envelope. 
Ice features in the Class I object envelopes differ somewhat from those seen around such
objects in other star-forming regions. While the same set of features is present, their
amplitude is lower, per unit silicate dust absorption, and the wavelengths are slightly shifted.
Both of these effects suggest the Class I envelopes are warmer than in Class I sources in other
star-forming regions, perhaps due to the warmer gas temperature in the star-forming material
in the globule IC 1396A.
The somewhat different physical conditions of the Class I envelopes in this globule
will affect the chemistry at the time when planet formation is just beginning.
Because the Sun is thought to have formed near an O star, 
and most nearby star-forming regions lack an O star, 
astronomical searches for the physical conditions of the early solar 
nebula would be best conducted in regions like IC 1396.

The stars photometrically classified as Class II are confirmed as such based on the
presence of optical spectra with bright H$\alpha$ emission lines and mid-infrared
spectra with silicate features in emission. Most of the disk spectra are similar to
other commonly-seen T Tauri star spectra, dominated by amorphous silicates. But the
star IC 1396A:$\theta$ has a particularly structured silicate feature with two 
`horns' at 9.1 and 11.3 $\mu$m. The 11.3 $\mu$m horn can be explained by crystalline olivine.
The origin of the 9.1 $\mu$m horn cannot be uniquely identified, with reasonably good
fits provided by quartz (SiO$_{2}$), a phyllosilicate (montmorillonite), or 
$>12$ $\mu$m radius amorphous olivine grains. Similar spectra are evident in other
wide samples of T Tauri stars.
Such spectra show that the planet-building material around at least some stars is 
dominated by crystalline silicates and is significantly different from interstellar
silicates.

The star formation history in IC 1396A cannot be described as a single ``burst'' of 
star formation: Class I and II sources in and near the globule span a range of ages 
and masses in a small region (2 pc across).
It appears that the earliest star formation
in IC 1396 occurred approximately 5 Myr ago,
based on the ages of the Class II sources and the central O6 star.
Some stars ahead of the present-day bright rim of IC 1396A 
have become exposed by the recession
of the globule and by their own outflows.
Recently, $< 1$ Myr ago, a powerful wind from the young 
AeBe star LkH$\alpha$ 349a blew
a bubble through the center of the globule.
The star formation in IC 1396A we observe now
as Class I sources occurred within the last 0.2 Myr,
based on the ages of the Class I sources.

The most recent star formation occurs preferentially behind the center of the globule,
just inside bright rims.
The specific trigger for the most recent star formation may have included
not only  pressurization of the globule by the O star exciting IC 1396,
but also 
the pressure of outflows and jets from the earlier generation of stars
and in particular those with powerful winds like LkH$\alpha$ 349a.
The young stars within IC 1396A appear to be reshaping the
globule from the inside, clearing `compartments' within the globule.
Gradually, the globule is being fragmented due to the exponentially
increasing number of stars (if indeed they are sequentially triggered).
The mass contained in walls of the compartments will eventually become
small enough, and their overlapping structure chaotic enough, that 
further star formation in the globule will end. The limiting factor
in the star formation efficiency of this one globule will be the
dynamic effect of energy input from the stars formed within it.

\acknowledgements  
WTR gratefully acknowledges exceptionally valuable discussions with Bruce 
Elmegreen regarding the dynamics of the globule.
Geoff Blake is thanked for assistance in obtaining the
Keck/NIRSPEC $L$-band spectra.

This work is based in part on observations made with the {\it Spitzer Space
Telescope}, which is operated by the Jet Propulsion Laboratory, California
Institute of Technology under NASA contract 1407. 

This work is based in part on observations obtained at the Hale Telescope,
Palomar Observatory as part of a continuing collaboration between the 
California Institute of Technology, NASA/JPL, and Cornell University.

Some of the data presented herein were obtained at the W.M. Keck
Observatory, which is operated as a scientific partnership among the
California Institute of Technology, the University of California and
the National Aeronautics and Space Administration. The Observatory was
made possible by the generous financial support of the W.M. Keck
Foundation.

This publication makes use of data products from the Two Micron All Sky Survey, which is a joint project of the University of Massachusetts and the Infrared Processing and Analysis Center/California Institute of Technology, funded by the National Aeronautics and Space Administration and the National Science Foundation.

\clearpage

\begin{deluxetable}{lccccc}
\tablewidth{0pt}
\tablecaption{Source list and observing log\label{opttab}}
\tablehead{
\colhead{Name\tablenotemark{a}} & \colhead{RA DEC}& \colhead{Class\tablenotemark{b}} & \colhead{IRS\tablenotemark{c}} & \colhead{DBSP\tablenotemark{d}} & \colhead{GBT\tablenotemark{e}} 
}
\startdata
IC1396A:$\alpha$ & 21$^h$36$^m$46.6$^s$ +57$^\circ$29$'$38$''$   & I    & SL+LL+H & o  & Y    \\
IC1396A:$\beta$  & 21$^h$36$^m$55.2$^s$ +57$^\circ$30$'$30$''$   & II   & SL+LL & o  & Y    \\
IC1396A:$\gamma$ & 21$^h$36$^m$08.0$^s$ +57$^\circ$26$'$37$''$   & I/0  & SL+LL+H & o  & Y    \\
IC1396A:$\delta$ & 21$^h$36$^m$19.3$^s$ +57$^\circ$28$'$38$''$   & I/0  & SL+LL & o  & Y    \\
IC1396A:$\epsilon$&21$^h$36$^m$18.4$^s$ +57$^\circ$28$'$30$''$   & I/0  & SL+LL & o  & Y    \\
IC1396A:$\eta$   & 21$^h$36$^m$57.8$^s$ +57$^\circ$30$'$55$''$   & I    & SL+LL & o  & Y    \\
IC1396A:$\theta$ & 21$^h$36$^m$39.2$^s$ +57$^\circ$29$'$53$''$   & II   & SL+LL+H & Y  & Y    \\ 
IC1396A:$\iota$  & 21$^h$36$^m$36.8$^s$ +57$^\circ$31$'$32$''$   & II   &  ... &...& Y    \\
IC1396A:$\kappa$ & 21$^h$36$^m$56.4$^s$ +57$^\circ$31$'$51$''$   & II   &  ... &...& Y    \\ 
IC1396A:$\lambda$& 21$^h$36$^m$54.8$^s$ +57$^\circ$30$'$00$''$   & I    & SL+LL & o  & Y    \\
IC1396A:$\mu$    & 21$^h$37$^m$02.9$^s$ +57$^\circ$30$'$48$''$   & I    &  ... &...& Y    \\
IC1396A:$\nu$    & 21$^h$36$^m$57.9$^s$ +57$^\circ$29$'$11$''$   & I/II &  ... &...& Y    \\
IC1396A:$\xi$    & 21$^h$35$^m$57.9$^s$ +57$^\circ$29$'$11$''$   & I/II &  ... &...& Y    \\
IC1396A:$\zeta$  & 21$^h$37$^m$02.3$^s$ +57$^\circ$31$'$15$''$   & I    &  ... &...& Y    \\
LkH$\alpha$349c  & 21$^h$36$^m$49.5$^s$ +57$^\circ$31$'$22$''$   & II   & SL+LL+H & Y  & ... \\ 
LkH$\alpha$349a  & 21$^h$36$^m$50.7$^s$ +57$^\circ$31$'$10$''$   & III   & SL    & Y  & ... \\ 
Tr37 11-2146     & 21$^h$36$^m$57.6$^s$ +57$^\circ$27$'$32$''$   & II   & SL+LL & Y  & ... \\ 
Tr37 11-2037     & 21$^h$37$^m$07.0$^s$ +57$^\circ$27$'$01$''$   & II   & SL+LL & Y  & ... \\ 
Tr37 72-1427     & 21$^h$35$^m$16.3$^s$ +57$^\circ$28$'$22$''$   & II   & SL+LL & Y  & ... \\ 
21361664+5728404 & 21$^h$36$^m$16.6$^s$ +57$^\circ$28$'$40$''$   & II   & SL    & Y  & ... \\ 
21364964+5722270 & 21$^h$36$^m$49.7$^s$ +57$^\circ$22$'$27$''$   & II   & SL+LL & Y  & ... \\ 
Tr37 72-875      & 21$^h$35$^m$49.8$^s$ +57$^\circ$24$'$04$''$   & II   & SL+LL & Y  & ... \\ 
21364398+5729287 & 21$^h$36$^m$43.9$^s$ +57$^\circ$29$'$28$''$   & II   & SL+LL & Y  & ... \\ 
21362507+5727502 & 21$^h$36$^m$25.1$^s$ +57$^\circ$27$'$50$''$   & II   & SL+LL & Y  & ... \\ 
\enddata
\tablenotetext{a}{Designations from IC1396A:x from Paper I, Simbad: [RRY2004] x; designations from \citep{sicilia05}, Simbad: [SHB2004] Trumpler 37 x-y; numeric designations from 2MASS All-Sky Point Source Catalog}
\tablenotetext{b}{Classification based on mid-infrared colors (Paper I).}
\tablenotetext{c}{Observed with the Spitzer/Infrred Spectrograph; SL=5.2-14 $\mu$m, LL=14-35 $\mu$m, H=high-res (10-35 $\mu$m).}
\tablenotetext{d}{Observed with the Palomar 5-m/Double Spectrograph; Y=good spectrum, o=observed but not detected (no optical counterpart).}
\tablenotetext{e}{Observed with the NRAO Green Bank Telescope for 22 GHz H$_{2}$O masers.}
\end{deluxetable}  


\begin{deluxetable}{lrrrrrrrrrrr}
\tablewidth{0pt}
\tablecaption{Brightness (magnitudes) of observed sources\label{magtab}}
\tablehead{
\colhead{Name}  & \colhead{B}  & \colhead{V}  & \colhead{R}  & \colhead{J}  & \colhead{H}  & \colhead{K$_s$}  & \colhead{[3.6]}  & \colhead{[4.5]}  & \colhead{[5.8]}  & \colhead{[8.0]}  & \colhead{[24]}
}
\startdata
IC1396A:$\alpha$   &  ...   &  ...   &  ...   &  17.32 &  15.31 &  13.51  &  9.858  &  8.44  & 7.19  &  6.28  &  2.01  \\ 
IC1396A:$\beta$    &  ...   &  ...   &  ...   &  13.83 &  12.47 &  11.29  &  9.832  &  9.97  & 8.75  &  8.03  &  5.04  \\
IC1396A:$\gamma$   &  ...   &  ...   &  ...   &  19.76 &  ...   &  13.95  &  10.56  &  9.22  & 7.84  &  6.80  &  2.24  \\
IC1396A:$\delta$   &  ...   &  ...   &  ...   &  $>20$ &  ...   &  15.76  &  12.66  &  10.9  & 9.51  &  8.44  &  3.73  \\
IC1396A:$\epsilon$ &  ...   &  ...   &  ...   &  $>20$ &  ...   &  16.95  &  12.57  &  10.9  & 9.16  &  8.55  &  4.27  \\
IC1396A:$\zeta$    &  ...   & ...    &  ...   &  ...   &  16.48  & 13.62  &  10.84  &  9.86  & 9.02 &   8.84 &  4.48 \\
IC1396A:$\eta$     &  ...   &  ...   &  ...   &  17.08 &  15.41 &  14.05  &  12.40  & 11.36  & 10.6  &  9.15  &  4.93  \\
IC1396A:$\lambda$  &  ...   &  ...   &  ...   &  17.89 &  15.72 &  14.06  &  11.92  & 11.55  & 10.50  &  9.98  &  5.91  \\
IC1396A:$\theta$   &  ...   & ...    &  17.0  &  11.73 &  10.21 &   9.50  &  8.414  &  7.85  & 7.21  &  6.53  &  4.07  \\
LkH$\alpha$349c    &  17.0  &  ...   &  12.4  &  10.59 &  10.81 &  10.13  &  9.600  &  9.23  & 8.77  &  8.65  &  4.94  \\
LkH$\alpha$349a    &  15.2  &  ...   &  11.4  &  11.71 &   8.95 &   8.62  &  8.440  &  8.44  & 8.25  &  8.14  &  ...   \\
Tr37 11-2146 	  &  17.1  & 16.92  &  15.6  &  12.07 &  11.52 &  10.82  &  9.932  &  9.69  & 9.13  &  8.34  &  4.89  \\
Tr37 11-2037       &  16.3  & 16.03  &  15.1  &  12.58 &  11.72 &  11.29  &  10.87  & 10.59  & 10.2  &  9.51  &  4.49  \\
Tr37 72-1427       &   ...  & 18.08  &  ...   &  14.03 &  13.07 &  12.67  &  12.08  & 11.89  & 11.4  &  10.8  &  6.81  \\
21361664+5728404	  &   ...  &  ...   &  ...   &  17.02 &  15.06 &  13.98  &  12.29  & 11.57  & 10.8  &  10.0  &  7.22  \\
21364964+5722270   &  16.8  &  ...   &  14.9  &  12.84 &  12.07 &  11.79  &  11.38  & 11.51  & 10.9  &  10.3  &  5.73  \\
Tr37 72-875        &  ...   &  18.37 &  ...   &  14.17 &  13.04 &  12.51  &  11.60  & 11.31  & 10.7  &  10.0  &  6.73  \\
21364398+5729287   &  ...   &  ...   &  ...   &  14.34 &  13.07 &  12.38  &  11.64  & 11.24  & 10.8  &  10.6  &  6.35  \\
21362507+5727502	  &  ...   &  ...   &  ...   &  14.84 &  13.94 &  13.39  &  12.39  & 11.96  & 11.5  &  10.4  &  7.58  \\
\enddata
\end{deluxetable}

\begin{deluxetable}{lrlrrrr}
\tablewidth{0pt}
\tablecaption{Derived properties from optical spectra\tablenotemark{a}\label{optmore}}
\tablehead{
\colhead{name}  & $10^{14}I($H$\alpha)$ & H$\beta$/H$\alpha$ & $A_{V}^{\beta/\alpha}$ & $A_{V}^{SED}$ & SpType & Age
}
\startdata
IC1396A:$\alpha$ & 1.12     & 0.134 & 3.6  &  10.1  & ...  & ...   \\
IC1396A:$\beta$  & 0.86     & 0.12  & 4.1  &  14.0  & F0\tablenotemark{b}  & ...   \\
IC1396A:$\gamma$ & 1.05     & 0.19  & 2.3  & (1.5)  & ...  & ... \\  
IC1396A:$\delta$ & 0.87     & 0.16  & 3.0  &  4.2  & ...  & ...  \\  
IC1396A:$\theta$ & 1.28      & 0.071 & 6.0  &  3.8   & A2\tablenotemark{b}  & ...   \\
LkH$\alpha$ 349c & 2.85      & 0.092 & 5.0  &  3.6   & G9\tablenotemark{b}  & ...   \\ 
Tr37 11-2146     & 4.71      & 0.10  & 3.4  &  1.5   & K6.0 & 0.9 \\
Tr37 11-2037     & 11.9      & 0.14  & 3.3  &  1.5   & K4.5 & 2.5 \\
Tr37 72-1427     & 4.8       & 0.15  & 3.1  &  1.5   & M1.0 & 2.2 \\   
21361664+5728404 & 1.13      & 0.177 & 2.5  &  3.6   & ...  & ...  \\
21364964+5722270 & 5.02      & 0.066 & 6.2  &  1.5   & G5\tablenotemark{b}  & ... \\
Tr37 72-875      & 2.75      & 0.145 & 3.3  &  2.4   & M0.5 & 8.4 \\
21364398+5729287 & 2.41      & 0.071 & 6.0  &  7.1   & ...  & ... \\
21362507+5727502 & 3.26      & 0.17  & 2.7  &  1.7   & M0.0 & ... \\
\enddata
\tablenotetext{a}{H$\alpha$ and H$\beta$ fluxes were measured for each source.
For the Class I objects, the spectra were extracted at the position of the infrared source and
integrated over the appropriate wavelengths, but
no optical counterparts were evident (so the lines are most likely nebular).
Spectral types are from \citet{sicilia05}}.
\tablenotetext{b}{approximate spectral type based on luminosity}
\end{deluxetable}  


\begin{deluxetable}{cccc}
\tablewidth{0pt}
\tablecaption{22 GHz Water masers in IC 1396A\label{masertab}}
\tablehead{
 \colhead{Date} & \colhead{$F_\nu^{peak}$} & \colhead{$F_{H_2 0}$} & \colhead{$L_{H_2 0}$}\\
&                    \colhead{(Jy)}           & \colhead{(Jy~km~s$^{-1}$)} &  \colhead{($10^{-9} L_\odot$)}
}
\startdata
 2006 Aug 07  &   0.030     &          0.02     &       0.2 \\ 
 2006 Sep 28  &   0.042     &          0.06     &       0.7  \\ 
 2006 Nov 18  &   0.032     &          0.02     &       0.2 \\ 
 2007 Jan 24  &   0.29      &          0.28     &       3.7\\ 
\enddata
\end{deluxetable}


\begin{deluxetable}{lrrrrrr}
\tablewidth{0pt}
\tablecaption{Mid-infrared spectral properties of protostars in 
IC 1396A\label{protoproptab}\tablenotemark{a}}
\tablehead{
\colhead{Source} & \colhead{[8]-[24]} & \colhead{[3.6]-[5.8]} & \colhead{$\alpha_{5-8}$} & 
\colhead{$\Delta F_{10}/F$ } 
 & \colhead{$\Delta F_{15.2}/F$} & \colhead{$\Delta F_{6.0}/F$}
}
\startdata
\cutinhead{Class I Sources}
IC1396A:$\alpha$ &  4.27 & 2.66   & -0.32 & -0.64 &  -0.13  &  -0.08\\ 
IC1396A:$\gamma$ &  4.55 & 2.72   & +0.19 & -0.70 &  -0.10  &  -0.06\\
IC1396A:$\delta$ &  4.70 & 3.15   & +0.16 & -0.79 &  -0.18  & ...\\
IC1396A:$\epsilon$& 4.27 & 3.41   & -0.04 & -1.00 &  -0.25  &  -0.17 \\
IC1396A:$\eta$   &  4.22 & 1.75   & +0.24 & -0.31 &  	... 	 &  ...\\
IC1396A:$\lambda$&  3.57 & 1.26   & -0.12 & -0.15 &  	... 	 &  ...\\
\cutinhead{Class II Sources}                                     
IC1396A:$\beta$  &  2.99 & 1.07   &-0.26  & ...   && \\
IC1396A:$\theta$ &  2.45 & 1.19   &-0.22  & 0.208 && \\
LkH$\alpha$349a  &  1.62 & 0.82   &-0.30  & ...  	&&\\
LkH$\alpha$349c  &  ...  & 0.18   &-0.49  & ...	&&\\
Tr37 11-2146     &  3.44 & 0.79   &-0.38  & 0.817 && \\ 
Tr37 11-2037     &  5.01 & 0.63   &-0.31  & 0.507 &&\\
Tr37 72-1427     &  4.08 & 0.58   &-0.31  & ...   &&\\
21364964+5722270 &  4.61 & 0.38   &-1.97  & 1.94  &&\\
Tr37 72-875      &  3.31 & 0.87   &-0.03  & ... 	&&\\
21364398+5729287 &  4.32 & 0.78   &-0.95  & ... 	&&\\
21362507+5727502 &  2.83 & 0.88   &-0.15  & 0.625 &&\\
\enddata
\tablenotetext{a}{[a]-[b] is the color (magnitude difference) between bands a and b; 
$\alpha_{5-8}$ is the spectral slope $F_\nu\propto \nu^{\alpha}$ from 5 to 8 $\mu$m;
$\Delta F_{\lambda}/F$ is the amplitude emission (or absorption, if negative) feature, 
centered on wavelength $\lambda$ ($\mu$m), relative to the local continuum.}
\end{deluxetable}

\begin{deluxetable}{lrrr}
\tablewidth{0pt}
\tabletypesize{\scriptsize}
\tablecolumns{4}
\tablewidth{0pc}
\tablecaption{Ice Column Densities\label{t:colden}}
\tablehead{
\colhead{Source}& \colhead{$N$(H$_2$O)}     & \colhead{$N$(CO$_2$)} & \colhead{$\tau_{9.7}$} \\
\colhead{      }& \colhead{$10^{18}~{\rm cm^{-2}}$ } & \colhead{\% H$_2$O  } & \colhead{} \\}
\startdata
IC 1396A $\alpha$   &  1.6 (0.2)             &  14 & 1.12 (0.11)\\
IC 1396A $\gamma$   &  1.2 (0.2)             &  34 & 1.41 (0.14)\\
IC 1396A $\delta$   &  4.3 (0.8)             &  23 & 1.96 (0.20)\\
IC 1396A $\epsilon$ &  6.8 (1.7)             &  26 & 9.64 (1.00)\\
\enddata
\end{deluxetable}

\def\extra{
\clearpage
\begin{deluxetable}{lcc}
\tablewidth{0pt}
\tablecaption{Ice absorption feature strengths\label{icetab}}
\tablehead{
\colhead{source} & \colhead{$\Delta F_{15.2}/F_{15.2}$} & \colhead{$\Delta F_{6}/F_{6}$}
}
\startdata
alpha  &    -0.110 &    ...	\\	   
delta  &  	... 	  &  ...\\
epsilon&    -0.259 &  -0.281\\		 
eta    &  	... 	  &  ...\\
gamma  &    -0.111 &    ...\\
lambda &  	... 	  &  ... \\
\enddata
\end{deluxetable}
}

\begin{deluxetable}{lrrrrl}
\tablewidth{0pt}
\tablecaption{Empirical model for Class I protostars in IC 1396A\label{protomodeltab}}
\tablehead{
\colhead{} & \colhead{IC1396A:$\gamma$} & \colhead{$\alpha$} & \colhead{$\delta$} & \colhead{$\epsilon$}
& \colhead{Units} }
\startdata
$\tau_{sil}$        &  1.7 &  1.5 &  2.1 &  4.0 &   \\
$A_{V}^{2}$         & 24.5 & 21.6 & 30.2 & 57.6 &   \\
$A_{V}^{3}$         &  0.5 &  0.9 &  2.0 &  3.4 &   \\
$R_{acc}$           &   27 &   24 &  209 &  279 & $R_\odot$\\
$R_{2}$             &  3.3 &  3.3 &  6.8 &  5.9 & AU\\
$R_{3}$             &  9.5 &  1.8 &  1.6 &  1.7 & $10^{3}$ AU  \\
$n_{2}$             &  0.9 &  0.8 &  0.6 &  1.2 & $10^{9}$ cm$^{-3}$ \\
$n_{3}$             &  0.6 &  6.4 & 16.3 & 24.8 & $10^{4}$ cm$^{-3}$   \\
$\dot{M}$           &  2.3 &  2.0 &  4.1 &  7.3 & $10^{-6} M_{\odot}$~yr$^{-1}$\\
$L_{1}$             &  2.7 &  2.7 &  0.6 &  0.8 & $L_{\odot}$  \\
$L_{tot}$           & 12.3 &  5.0 &  1.8 &  2.6 & $L_{\odot}$ \\
\enddata
\end{deluxetable}  


\begin{deluxetable}{lrrrrrrl}
\tablewidth{0pt}
\tablecaption{Radiative transfer models for Class I protostars in IC 1396A\label{whitneytab}}
\tablehead{
\colhead{} & \colhead{IC1396A:$\gamma$} & \colhead{$\alpha$} & \colhead{$\delta$} & \colhead{$\epsilon$}
& \colhead{$\eta$} & \colhead{$\zeta$} & \colhead{Units} }
\startdata	             
Model\tablenotemark{a}
                 & 3016360 & 3000224 & 3017673 & 3016207 & 3003269 & 3013344 & \\
$\chi^{2}_{\nu}$ & 5.09    &  1.41   & 3.5     & 6.8      & 2.24    & 1.6     & \\ 
$\dot{M}$        & 91.4    & 2.39    & 2.55    & 0.76     & 19.3    & 27.8    & $10^{-6} M_{\odot}$~yr$^{-1}$\\
$R_{*}$          & 10.0    & 6.13    & 4.94    & 4.10     & 3.53    & 3.39    & $R_{\odot}$\\
$M_{*}$          & 1.88    & 0.38    & 0.15    & 0.11     & 0.20    & 0.43    & $M_{\odot}$\\
$T_{*}$          & 4321    & 3503    & 2839    & 2581     & 3062    & 3650    &  K\\
Age($*$)         & 11.1    & 0.14    & 0.16    & 0.18     & 4.00    & 19.3    &  $10^4$ yr \\
$L_{tot}$        & 36.2    & 5.37    & 1.58    & 0.67     & 0.99    & 2.68    & $L_{\odot}$\\
$i$              & 69.5    & 31.8    & 41.4    & 56.6     & 41.4    & 63.3    & $^\circ$\\
$A_{V}^{int}$    & 65.2    & 62.9    & 64.3    & 73.6     & 26.5     & 30.4    & mag (to central star)\\
$A_{V}$          & 7.4     & 1.4     & 8.0     & 7.4      & 4.4     & 4.4     & mag (line of sight)
\enddata
\tablenotetext{a}{Model identification from \citet{robitaille}.}
\end{deluxetable}


\begin{deluxetable}{lrrrrrrl}
\tablewidth{0pt}
\tabletypesize{\scriptsize}
\tablecaption{Radiative transfer models for Class II protostars in IC 1396A\label{whitneytabII}}
\tablehead{
  		        &  \colhead{21364964+5722270} &  \colhead{IC1396A:$\beta$} & \colhead{LkH$\alpha$ 349c} & 
\colhead{LkH$\alpha$ 349a} &
\colhead{IC1396A:$\theta$} & \colhead{Tr37 11-2037} & \colhead{Units}
}
\startdata	
Model\tablenotemark{a}
           & 3004629 & 3015197 & 3002727 & 3005288 & 3017958 & 3007956 &\\
$\chi^2_{\nu}$  & 3.9     & 0.6     & 8.7     & 6.1     & 5.6     & 1.1     &\\
$M_{disk}$      & 0.009   & 31.9    & 25.9    & 0.67    & 0.00007 & 0.26    &  $10^{-3} M_{\odot}$  \\
$\dot{M}_{disk}$& 0.007   & 68.2    & 70.9    & 0.01    & 0.002   & 1.53    & $10^{-9} M_{\odot}$~yr$^{-1}$ \\
$R_{*}$         & 1.77    & 2.34    & 4.41    & 3.63    & 2.67    & 2.24    &$R_{\odot}$\\
$M_{*}$         & 1.50    & 0.80    & 0.70    & 2.28    & 4.99    & 0.80    &$M_{\odot}$   \\
$T_{*}$         & 5247    & 4123    & 3988    & 4829    & 16,450  & 4124    & K\\
Age($*$)        & 91.6    & 10.1    & 4.1     & 12.4    & 31.8    & 11.7    & $10^5$ yr \\
$L_{tot}$       & 2.1     & 7.59    & 4.73    & 6.43    & 46.9    & 1.31    & $L_{\odot}$   \\
$i$             & 41.4    & 69.5    & 69.5    & 81.4    & 41.4    & 18.2    & $^\circ$\\
$A_{V}^{int}$   & 0.002   & 0.016   & 0.85    & 0.003   & 0.001   & 0.011   & mag (to central star)\\
$A_{V}$         & 2.8     & 6.8     & 1.80    & 1.60    & 8.0     & 1.40    & mag (line of sight)
\enddata
\tablenotetext{a}{Model identification from \citet{robitaille}.}
\end{deluxetable}

\begin{deluxetable}{lrrrrrrcrr}
\tablewidth{0pt}
\tablecaption{Proplyd-like objects in IC 1396\label{proplydtab}}
\tablehead{
\colhead{Name} & \colhead{RA} & \colhead{Dec} & 
                        \colhead{O* Dist} & \colhead{O* PA} & \colhead{Tail PA} 
				                    & \colhead{$I_8$} & \colhead{$R_{hole}$}  & \colhead{$L_{tail}$}\\
& & &  \colhead{(pc)} & \colhead{($^\circ$)} &  \colhead{($^\circ$)} &  \colhead{(MJy sr$^{-1}$)} & \colhead{(AU)}& \colhead{(AU)}
}
\startdata
A & 21$^h$36$^m$25.2$^s$ & +57$^\circ$39$'$48$''$ & 5.0 & 207 & 196 & 1.4 & 2000 & 10500\\
B & 21$^h$36$^m$24.2$^s$ & +57$^\circ$39$'$22$''$ & 5.0 & 207 & 199 & 1.5 & 1900 &  9800\\
C & 21$^h$38$^m$13.7$^s$ & +57$^\circ$40$'$30$''$ & 2.6 & 242 & 240 & 2.4 & 3300 & 14000\\
D & 21$^h$39$^m$12.2$^s$ & +57$^\circ$24$'$07$''$ & 1.2 & 112 &  35 & 3.1 & 5600 & 74000\\
E & 21$^h$36$^m$19.0$^s$ & +57$^\circ$26$'$59$''$ & 4.7 & 174 & 161 & 5.5 & 1900 & 25000\\
\enddata
\end{deluxetable}  


\begin{deluxetable}{lrrrrrrrrr}
\tablewidth{0pt}
\tablecaption{Selected Infrared Sources Upstream from IC 1396A\label{rimfront}}
\tablehead{
\colhead{2MASS} & \colhead{[24]} & \colhead{[8.0]} & \colhead{[5.8]} & \colhead{[4.5]} & \colhead{[3.6]} & \colhead{K$_s$} & \colhead{H} & \colhead{J} & V
}
\startdata
21370936+5729483 & 6.70&  9.83 & 10.31 & 10.84 & 11.19 & 11.84 & 12.33 & 13.39 & 18.2 \\
21371054+5731124 & 7.35& 10.06 & 10.41 & 10.89 & 11.25 & 11.73 & 12.17 & 13.09 & 16.74\\
21370649+5732316 & 6.99&  9.91 & 10.44 & 10.69 & 11.19 & 11.91 & 12.33 & 13.32 & 17.85\\
21371172+573329  & 6.95& 10.78 & 10.73 & 10.88 & 11.06 & 10.90 & 11.12 & 11.94 & 17.4\\
21370802+5734094 & 7.66& 7.970 & ...   &  8.01 &  7.74 &  7.86 &  8.37 &  9.62 & 14.5\\
\enddata
\end{deluxetable}  

\end{document}